# Vision Foundation Models for Computed Tomography

**Authors |** Suraj Pai*[1,2,3], Ibrahim Hadzic*[1,2,3], Dennis Bontempi[1,2,3], Keno Bressem[4,5], Benjamin H. Kann[1,3], Andriy Fedorov[6], Raymond H Mak[1,3], Hugo JWL Aerts[1,2,3,6]

**Affiliations |** [1] Artificial Intelligence in Medicine (AIM) Program, Mass General Brigham, Harvard Medical School, Harvard Institutes of Medicine, 77 Avenue Louis Pasteur, Boston, MA 02115, United States of America; [2]Radiology and Nuclear Medicine, CARIM & GROW, Maastricht University, Universiteitssingel 40, 6229 ER Maastricht, The Netherlands; [3]Department of Radiation Oncology, Brigham and Women's Hospital, Dana-Farber Cancer Institute, Harvard Medical School, 75 Francis Street and 450 Brookline Avenue, Boston, MA 02115, USA; [4]Department of Diagnostic and Interventional Radiology, Technical University of Munich, School of Medicine and Health, Klinikum rechts der Isar, TUM University Hospital, Ismaninger Str. 22, 81675 Munich; [5]Department of Cardiovascular Radiology and Nuclear Medicine, Technical University of Munich, School of Medicine and Health, German Heart Center, TUM University Hospital, Lazarethstr. 36, 80636, Munich; [6]Department of Radiology, Brigham and Women's Hospital, Dana-Farber Cancer Institute, Harvard Medical School, 75 Francis Street and 450 Brookline Avenue, Boston, MA 02115, USA;

*These authors contributed equally

**Running title |** Vision Foundation Model for Computed Tomography

**Corresponding author |** Hugo Aerts, Ph.D., Artificial Intelligence in Medicine (AIM) Program, Mass General Brigham, Harvard Medical School, Harvard Institutes of Medicine – HIM 343, 77 Avenue Louis Pasteur, Boston, MA 02115, P - 617.525.7156, F - 617.582.6037, Email: haerts@bwh.harvard.edu

**Abstract | Foundation models (FMs) have shown transformative potential in radiology by performing diverse, complex tasks across imaging modalities. Here, we developed CT-FM, a large-scale 3D image-based pre-trained model designed explicitly for various radiological tasks. CT-FM was pre-trained using 148,000 computed tomography (CT) scans from the Imaging Data Commons through label-agnostic contrastive learning. We evaluated CT-FM across four categories of tasks, namely, whole-body and tumor segmentation, head CT triage, medical image retrieval, and semantic understanding, showing superior performance against state-of-the-art models. Beyond quantitative success, CT-FM demonstrated the ability to cluster regions anatomically and identify similar anatomical and structural concepts across scans. Furthermore, it remained robust across test-retest settings and indicated reasonable salient regions attached to its embeddings. This study demonstrates the value of large-scale medical imaging foundation models and by open-sourcing the model weights, code, and data, aims to support more adaptable, reliable, and interpretable AI solutions in radiology.**

## INTRODUCTION

Radiologists face a complex range of analytical tasks - from precise anatomical localization to disease characterization and monitoring. Traditional computational approaches have relied on separate specialized models for each task, such as different systems for cardiac imaging[1] and abdominal analysis[2]. This fragmentation extends to characterization tasks, where individual expert models are required for diagnosis, monitoring, segmentation, comparison, and prognosis[3]. Developing specialized models creates significant operational overhead, requiring substantial resources to create and maintain multiple systems for different analysis needs.

Foundation Models (FMs) provide a unified solution to these challenges through their ability to learn and adapt across multiple tasks. The emergence of large annotated and unannotated datasets has accelerated the development of radiological models that can adapt efficiently across various tasks. Over recent years, various foundation models integrating images and text have been proposed for radiological applications. MedVersa performs multiple vision-based tasks, including radiology report generation and anatomical segmentation[4]. Merlin offers 3D medical image interpretation using electronic health records and radiology reports[5]. HAI-DEF models provide strong performance on medical tasks from several different domains leveraging Med-Gemini [6,7]. Despite advancements in multimodal learning, particularly in joint vision and language models, dedicated vision-centric encoders have been lacking in the field. Although many state-of-the-art models rely on vision encoding pathways, they are jointly trained with other modalities resulting in learning a joint distribution of features. A vision-centric encoder pre-trained solely on vision data could better represent the distribution of visual representations and complement existing foundation models. Leveraging more robust and fine-grained vision-specific features can potentially lead to improved performance on tasks requiring detailed visual understanding and improving their adaptability to novel visual domains.

Radiology, a domain heavily reliant on visual processing, has benefited largely from advances in image-based self-supervised learning (SSL) as a component of developing unified foundation models. These SSL methods enable models to learn robust feature representations from large quantities of unlabeled medical images, effectively addressing the chronic challenge of limited annotated data in healthcare settings[8–10]. Several studies have demonstrated that SSL pre-training can lead to superior performance compared to solely supervised task-specific training, both in general[11] and radiological tasks[8,9,12,13]. Moreover, some studies show that SSL pre-training can outperform models trained on large-scale labeled medical datasets[14]. Radiological image-based pre-training, however, is largely dominated by 2D pre-training performed slice-wise on 3D datasets. While select studies use large-scale native 3D pre-training, the pre-training methods, such as video pre-training, are not focused on spatial semantics in 3D imaging data[7,15]. Recent work in the radiology image pre-training has, therefore, focused on supervised pre-training due to the suboptimality of general SSL pre-training methods when applied to 3D data [16–19].

In this study, we present a foundation model with a self-supervised pre-training design focused on leveraging large-scale unannotated 3D imaging data for radiological interpretation tasks. The foundation model is pre-trained on 148,000 CT scans from the Imaging Data Commons and validated on (i) whole-body CT segmentation of 117 labels, (ii) highly heterogenous tumor segmentation across different anatomies, (iii) head CT triage for several conditions, (iv) content-based image retrieval and (v) semantic understanding tasks. We showed that CT-FM quantitatively outperformed our compared baselines and state-of-the-art approaches. In addition to these quantitative evaluations, we explored the semantics of our embeddings through anatomical clustering and concept identification discovering that CT-FM is inherently more interpretable due to the localized representations it learns during pre-training. Our results demonstrate that CT-FM is a pre-trained model that excels in segmentation and classification tasks in low-data settings, with embeddings that facilitate the retrieval of full CT scans or those scans that contain specific structures. By open-sourcing the CT-FM model pipeline, our complete dataset, and the implementation code developed in a reproducible and transparent framework, we aim to drive further research and evaluation accelerating the development of CT interpretation models for a wide range of clinical use-cases.

## RESULTS

In this study, we developed a foundation model for computed tomography applications using a dataset of 148,000 CT scans from a publicly available database of cohorts representing different cancer types, their phenotypes and associated comorbidities. The foundation model was pre-trained through a task-agnostic self-supervised learning strategy tailored to the unique characteristics of medical imaging data. The pre-trained foundation model was then evaluated through (i) fine-tuning on several clinically relevant tasks and (ii) zero-shot evaluations, which demonstrate its learning of concepts. An overview of the study is shown in **Figure 1**.

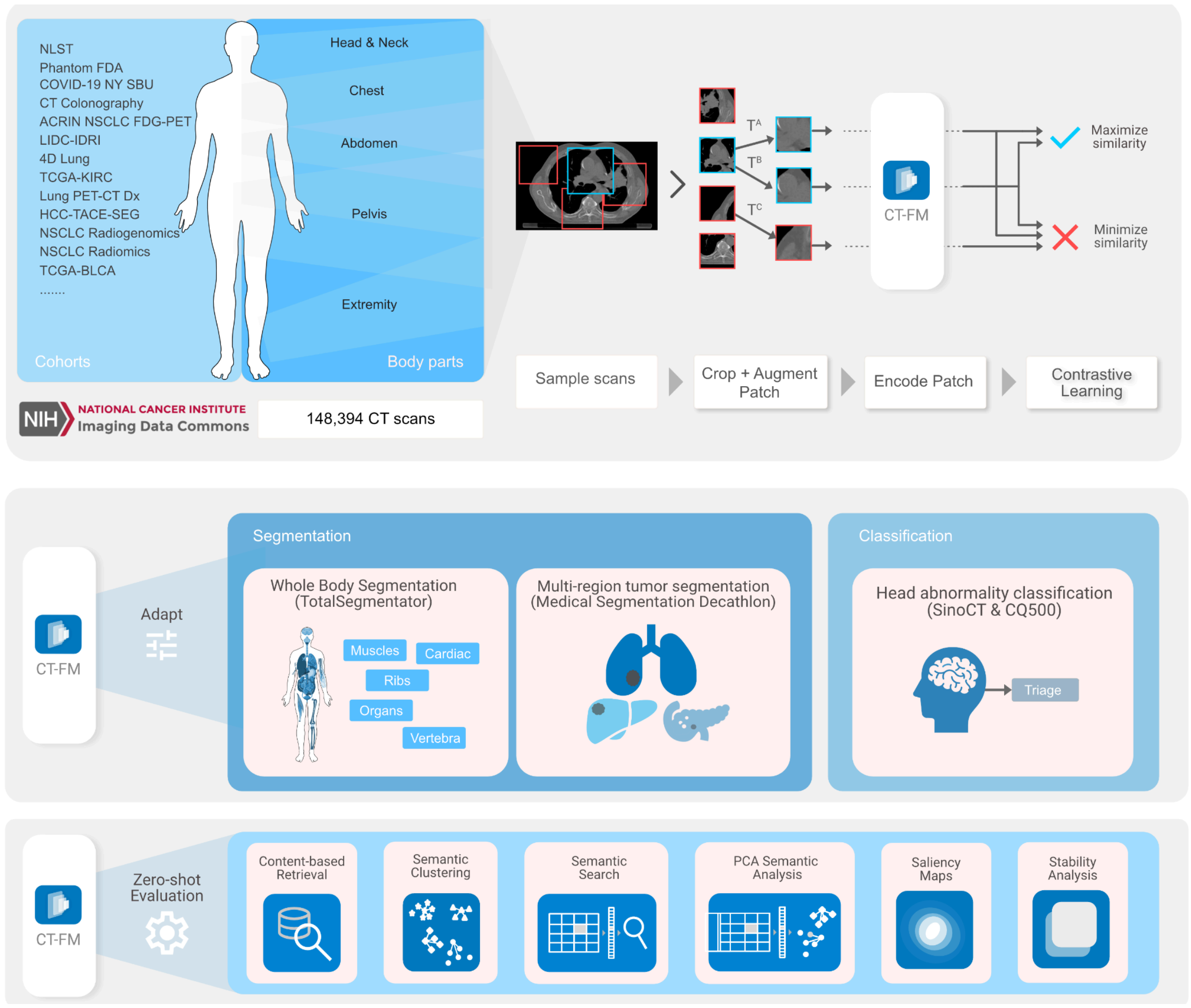

**Figure 1: Overview |** The CT-FM framework leverages contrastive pre-training on over 148,000 CT scans from the National Cancer Institute Imaging Data Commons to enable robust contrastive learning for medical image analysis. Adaptation tasks include whole-body segmentation, multi-region tumor segmentation, and head abnormality classification for triage. Zero-shot evaluation capabilities comprise of content-based retrieval, semantic search, clustering, PCA-based analysis, saliency mapping, and stability assessment, demonstrating CT-FM's versatility in segmentation, classification, and semantic understanding of CT images.

### Segmentation of 117 anatomical structures in whole-body CT scans

To assess the performance of our foundational model in dense segmentation, we compared performance of our foundational model against a model trained from scratch using an identical architecture but without foundational pre-training (serving as our architectural baseline), and a supervised segmentation foundational model, SuPREM [20]. The selection of SuPREM as a benchmark was justified by its robust performance in segmentation pre-training and its demonstrated superiority over previously established self-supervised methods. For this we used a dataset, consisting of 1228 scans annotated with 117 distinct expert defined anatomical structures[21]. Additionally, we compared previously reported results on the same split for models from other frameworks, namely, Auto3DSeg and VISTA3D [17]. The performance comparisons are detailed in **Figure 2**. Across the complete dataset (**Figure 2b**), our model exhibited a higher mean Dice coefficient (0.8981, 95% CI: 0.8959-0.9004) than the architectural baseline (0.8959, 95% CI: 0.8936-0.8982) and SuPREM (0.8695, 95% CI: 0.8668-0.8721). Our model also achieved better Dice scores than Auto3DSeg (0.882) and VISTA3D (0.893). We demonstrated that our baseline model also surpassed Merlin, achieving a mean Dice coefficient of 0.9017 (95% CI: 0.8885-0.9149) compared to 0.862. The foundational, pre-training further improved performance over the baseline (0.9058, 95% CI: 0.8929-0.9186) on the same split.

CT-FM was contrasted with the baseline and SuPREM models using specific label groups, including ribs, muscles, organs, cardiac structures, and vertebrae (see **Figure 2f**). CT-FM outperformed both the baseline and SuPREM across ribs, muscles, cardiac, and vertebra groups, while the baseline model showed better performance than CT-FM in the organ group. Considering individual labels (see **Figure 2d**), CT-FM showed better segmentation results in 73.5% of cases compared to the baseline model and 91.5% over SuPREM, indicating consistent improvements across most anatomical structures. Detailed comparison showing best improvements and performance on selected structures are shown in **Extended Data Figure 1.**

In the context of few-shot segmentation, which is gaining popularity in active learning annotation strategies such as MONAI Label[22], CT-FM demonstrated significantly improved performance at 5, 10, and 20 labels (see **Figure 2e**). We also show breakdown of few-shot segmentation across different structure groups in **Extended Data Figure 2** where we observe that a significant boost is offered using CT-FM for organs, muscles and cardiac groups. It is important to note that overall performance was on the lower end due to macro-averaging across 117 structures and high variability in smaller ROIs at these sample sizes. Nevertheless, these enhancements could offer advantages to end-users by reducing the need for corrections in the active learning process.

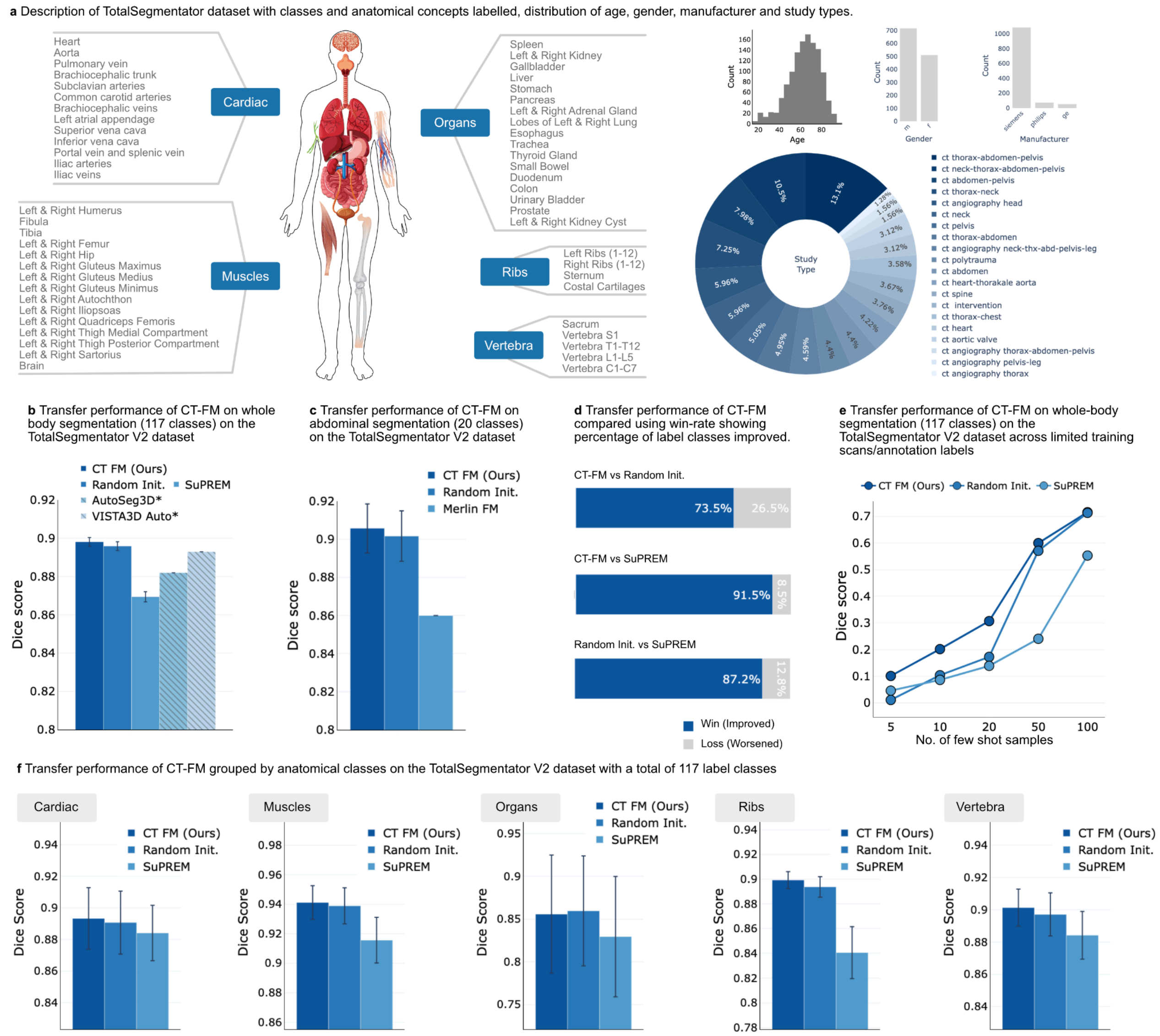

**Figure 2: Performance evaluation CT-FM on the TotalSegmentator dataset |** (a) Dataset overview: anatomical classes, demographics, and study types. (b-c) Transfer performance comparison (Dice score) on whole-body and abdominal segmentation. (d) CT-FM win-rate (%) against random initialization and SuPREM across label classes. (e) Few-shot learning performance with increasing training samples. (f) CT-FM performance across anatomical classes (cardiac, muscles, organs, ribs, vertebra).

## Segmentation of cancer across anatomical sites on CT scans

We evaluated CT-FM's performance on tumor segmentation tasks from the Medical Segmentation Decathlon dataset[23], focusing on hepatic, hepatic vessel adjacent, pancreatic and lung tumors. We specifically included a separate hepatic tumor task addressing the challenging case of heterogeneous tumors adjacent to hepatic vessels. To assess the value of CT-FM pre-training, we compared performance when integrating our weights into the established Auto3DSeg[24] framework versus standard initialization. Comparison between AutoSeg3D with and without CT-FM weights can be found in **Figure 3**.

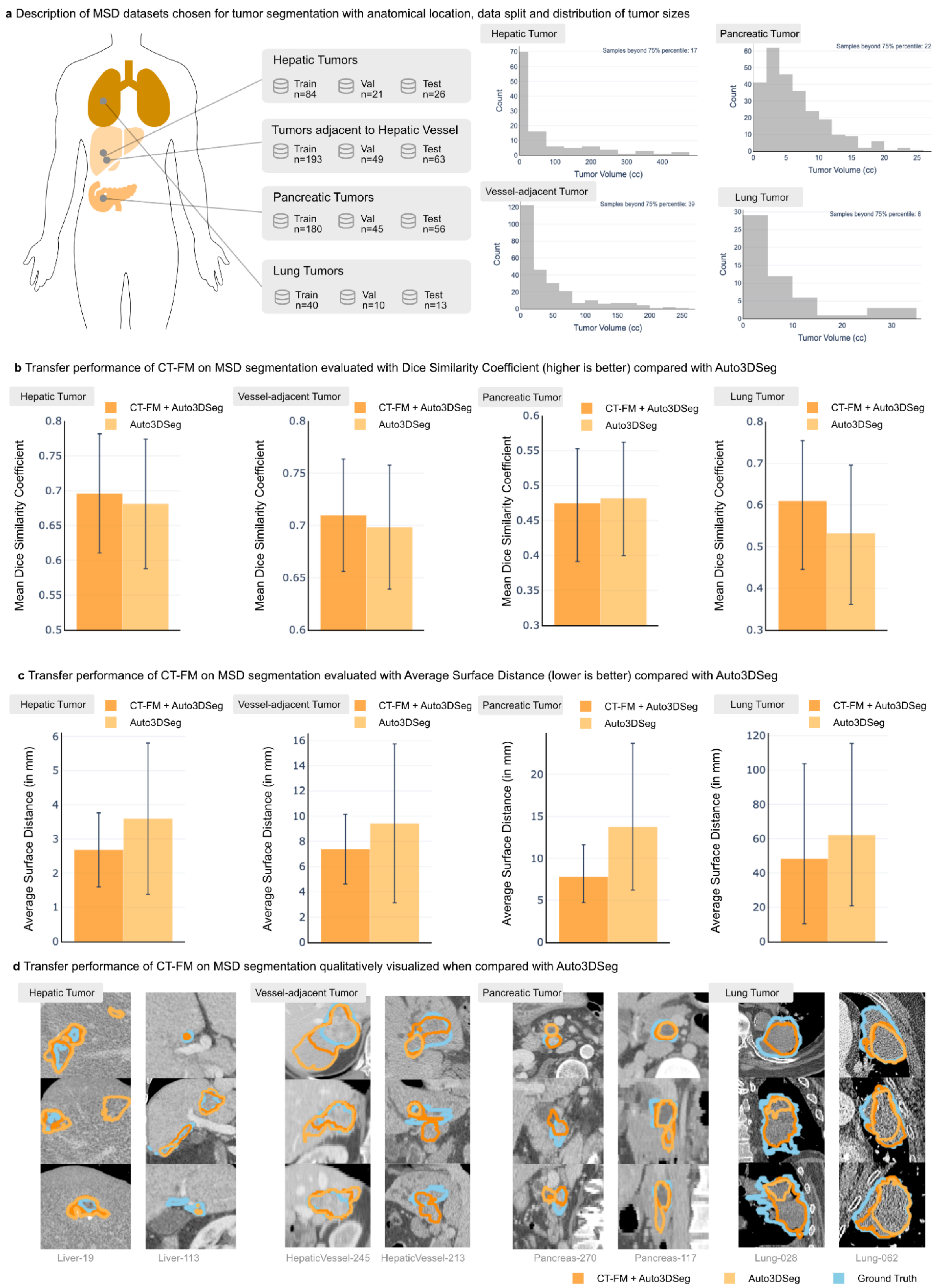

**Figure 3: Evaluation CT-FM on tumor segmentation using the Medical Segmentation Decathlon (MSD) dataset.** (a) MSD dataset overview: tumor types, anatomical locations, data splits, and tumor size distributions. (b-c) Quantitative comparison of CT-FM + Auto3DSeg vs. Auto3DSeg alone using Dice Similarity Coefficient (DSC) and Average Surface Distance (ASD) for hepatic, hepatic vessel-adjacent, pancreatic, and lung tumors. (d) Qualitative visual comparison of segmentation results for different tasks visualized through axial, coronal and sagittal views of sample scans.

For hepatic tumor segmentation, initializing Auto3DSeg with CT-FM weights led to improved performance, yielding a higher Dice score of 0.696 (95% CI: 0.612- 0.772) compared to 0.681 (95% CI: 0.590-0.759). The average surface distance (ASD) also decreased to 2.8 (95% CI: 1.7-3.7) mm from 3.6 (95% CI: 2.0- 6.0) mm. Considering hepatic tumors adjacent to the hepatic vessel, using CT-FM weights resulted in enhanced Dice scores of 0.709 (95% CI: 0.656-0.759) compared to 0.698 (95% CI: 0.637-0.753), and a reduction in ASD to 7.4 (95% CI: 4.9-10.2) mm from 9.4 (95% CI: 4.7-16.4) mm. In pancreatic tumor segmentation, although the Dice score showed no significant improvement with CT-FM weights (0.475, 95% CI: 0.395-0.549 vs. 0.482, 95% CI: 0.402-0.561), a noticeable improvement in ASD was observed, decreasing to 7.8 (95% CI: 4.8-11.5) mm from 13.8 (95% CI: 6.2-23.7) mm. Visual inspection confirmed that ASD was a more appropriate metric here, revealing fewer false positives and more true positives with CT-FM weights. Finally, when used for lung tumor segmentation, initialization with CT-FM weights improved both Dice scores (0.609, 95% CI: 0.445-0.754 vs. 0.532, 95% CI: 0.362-0.703) and average surface distance (ASD; 48.3, 95% CI: 10.8-103.4 vs. 62.1, 95% CI: 20.3-114.7) mm. Visualizations of the segmentations (**Figure 3d**) revealed that CT-FM weights reduced over-segmentation and improved the coverage of tumor extent. Across both the CT-FM and architectural baseline, we found that challenging tumor sites were missed which impacted the overall metric. We provide a case-wise breakdown of results and visualizations of missed ground truth in **Extended Data Figure 3**.

### Triage classification in head CT

Next, we evaluated CT-FM's performance in head CT triage using two complementary datasets. The SinoCT dataset includes over 9,000 head CT scans labeled by expert radiologists[25], while the CQ500 dataset consists of 491 scans with multi-category pathology annotations[26]. To allow direct comparability, CQ500's detailed categories were consolidated into a binary normal/abnormal classification to match the SinoCT dataset. Both datasets have an identical class distribution of 45% normal and 55% abnormal scans, enabling a balanced cross-dataset comparison. **Figure 4** shows the comparison between CT-FM and chosen baselines.

On the SinoCT dataset, the CT-FM model achieved an F1 score of 0.776 (95% CI: 0.750–0.802). The same architecture trained from scratch without pre-training yielded lower metrics, with an F1 score of 0.754 (95% CI: 0.727–0.781), but lower than the SuPREM model, which attained the highest F1 score of 0.798 (95% CI: 0.773–0.823). CT-FM reached an Area Under the Receiver Operating Characteristics Curve (AUC-ROC) of 0.836 (95% CI: 0.812–0.859). The baseline, non-pretrained model showed a lower AUC-ROC of 0.802 (95%

CI: 0.777–0.827), while SuPREM achieved the highest AUC-ROC of 0.868 (95% CI: 0.847–0.889).

To evaluate transferability, the models trained on SinoCT were tested in a zero-shot setting on the CQ500 dataset. Here, CT-FM achieved an F1 score of 0.754 (95% CI: 0.716–0.793) and AUC of 0.794 (95% CI: 0.758-0.829), surpassing both the non-pre-trained model (F1: 0.728, 95% CI: 0.689–0.767; AUC-ROC: 0.767, 95% CI: 0.729–0.804) and showing comparable performance to SuPREM (F1: 0.745, 95% CI: 0.706–0.783; AUC-ROC: 0.793, 95% CI: 0.757–0.829).

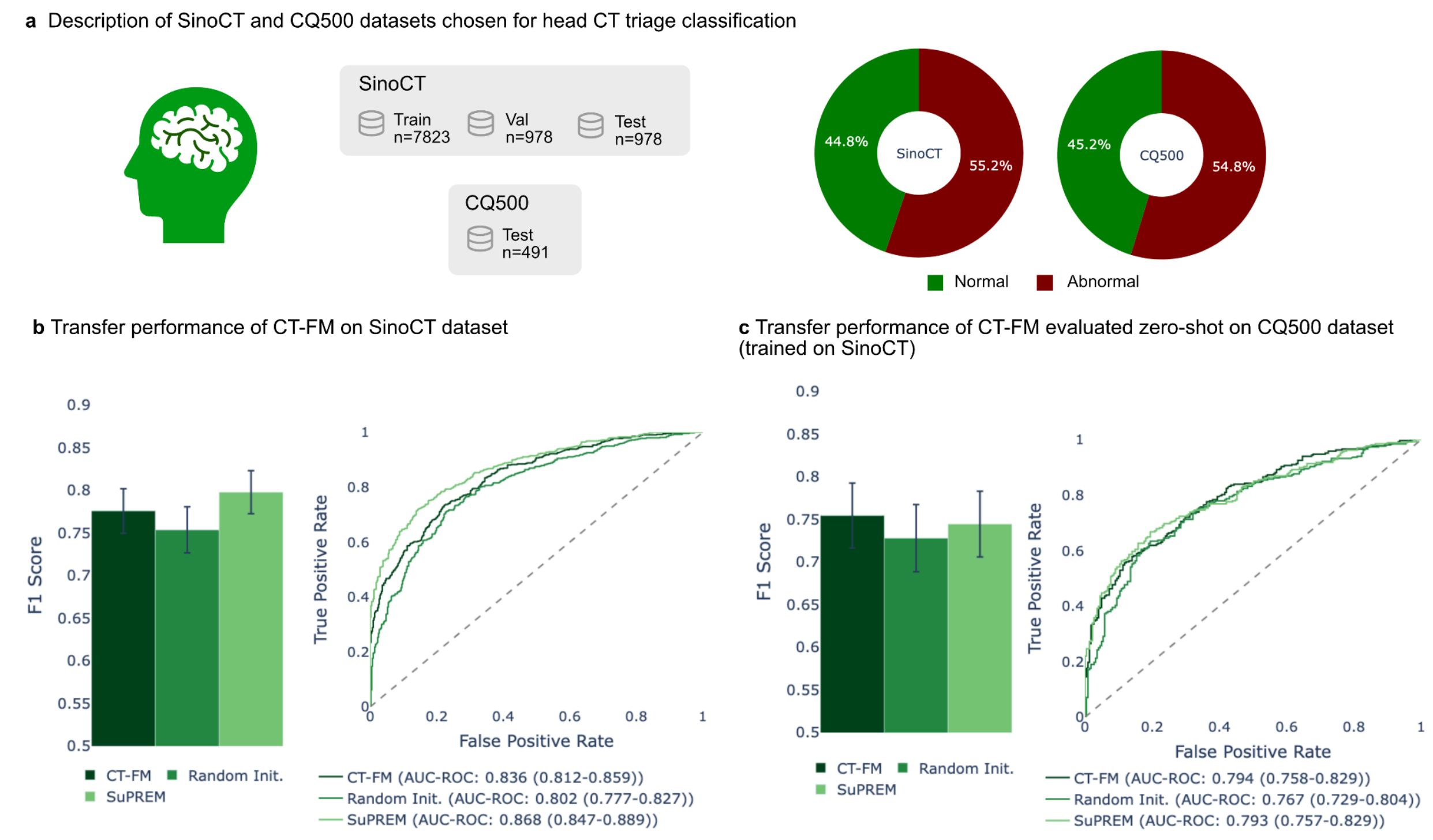

**Figure 4: Performance evaluation CT-FM on head CT triage classification using SinoCT and CQ500 datasets.** (a) Dataset descriptions: SinoCT (train/val/test splits) and CQ500 (test only), with normal/abnormal case distributions. (b) Transfer performance on SinoCT, measured by F1 score and ROC curve. CT-FM outperforms random initialization and shows comparable performance to SUPREM. (c) Zero-shot transfer performance on CQ500 (trained on SinoCT), measured by F1 score and ROC curve.

## Medical image retrieval

Embeddings of a CT foundation model should facilitate the search and retrieval of similar CT scans. To evaluate the quality of CT-FM-generated embeddings, we perform medical image retrieval tasks on two different datasets - OrganMNIST3D[27] and 3D-MIR [28], as shown in **Figure 5**. Embeddings in the test set are used to retrieve embeddings with the largest similarities in the train set. Posthoc labels are compared between the test set, and train set retrievals to analyze the retrieval performance. For the OrganMNIST3D, across different top match (top-k) percentages, CT-FM outperformed chosen baseline SuPREM on average precision (AP), hit rate (HR), and F1 score for retrieval of scans with a similar organ field-of-view (FOV). Across the most stringent criteria of k=3 matches, CT-FM shows improvements of AP=0.932, HR=0.968, and F1=0.944 compared to AP=0.923, HR=0.959 and F1=0.935 of SuPREM.

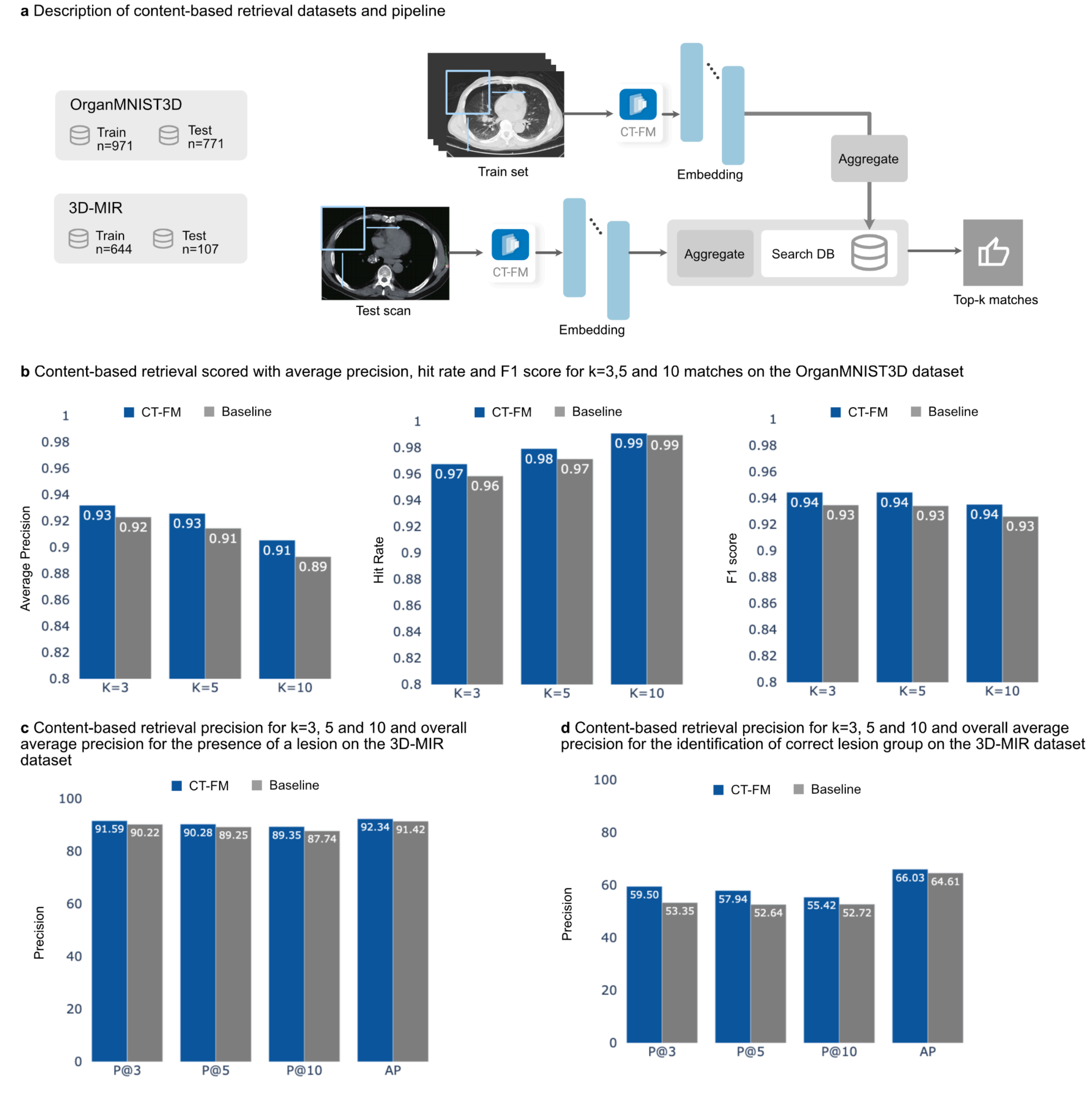

**Figure 5: Evaluation of CT-FM for content-based image retrieval (CBIR)**. (a) Datasets and pipeline: OrganMNIST3D and 3D-MIR datasets are used for training and testing, with a pipeline illustrating the embedding and search process. (b) CBIR performance on OrganMNIST3D using average precision, hit rate, and F1 score for k=3, 5, and 10. CT-FM outperforms the baseline across all metrics. CBIR precision on 3D-MIR for lesion presence detection (c) and lesion group identification (d) at k=3, 5, and 10, and overall average precision (AP).

3D-MIR defined a more complex task of retrieving scans with the presence of lesions as well as the identification of similar lesion groups. Compared to the baseline presented in the original paper, our method provided better precision across different match elements searched and averaged across all the matches. At k=3 matches, CT-FM retrieved scans containing lesions in the same anatomical site of the original scan with precision scores of 0.916 compared to 0.902 for the baseline. For identifying lesions with similar characteristics via lesion groups in the top 3 matches, CT-FM showed a precision of 0.595 compared to 0.534 for the dataset authors' baseline. Average precision across up to 10 matches showed improvements for CT-FM compared to baseline, with 0.923 vs. 0.914 for lesion identification and 0.660 over 0.646 for lesion grouping. We also present a detailed breakdown across cancer types in **Extended Data Figure 4**.

### Anatomical Clustering

To probe anatomical region awareness in the high-dimensional CT embedding, we select the OrganMNIST3D dataset, which contains resampled cubic volumes of 8 different anatomical regions with different FOV across several CT scans. We extracted embeddings from our CT-FM model and SuPREM on this dataset, which were then clustered through unsupervised tSNE. Post clustering, we assign colors to each point of their labeled anatomical site. As seen in **Figure 6a** and **Extended Data Figure 5**, CT-FM and SuPREM can attribute clusters to anatomical regions and distinguish between different anatomical regions.

### Semantic Concept Search

While our previous evaluation demonstrated CT-FM's capability in whole-scan retrieval (macro-level), meaningful embeddings in the CT latent space should also capture fine-grained correlations based on specific anatomical and conceptual structures (micro-level). To evaluate the relevance and interpretability of embeddings within localized regions, we selected scans of the same body part and assessed whether embeddings of specific anatomical structures (e.g., the heart) were more similar to each other than to embeddings of different structures using patch-based queries. Qualitative analysis showed that CT-FM has a superior anatomical concept consistency in the embedding space compared to SuPREM. As shown in **Figure 6b**, CT-FM can link concepts associated with heart, kidney, bowel, and cervical regions across scans containing these in their FOV compared to SuPREM, which fails to identify meaningful concepts at a micro-level. We also conduct a quantitative evaluation to compare CT-FM and SuPREM using a metric we develop called organ centroid distance (OCD) in **Figure 6c**. We observe that CT-FM provides a significantly lower distance of 5.61 ± 5.44 cm compared to 23.44 ± 7.06 cm of SuPREM. Moreover, in the case of CT-FM, 97.7% of the matched embedding patches were located within the same structure (heart) in the target scan, compared to 81.9% in the case of SuPREM. In **Extended Data Figure 6**, we also compare our pre-training strategy with other pre-training strategies and show that our chosen pre-training strategy leads to the development of this property.

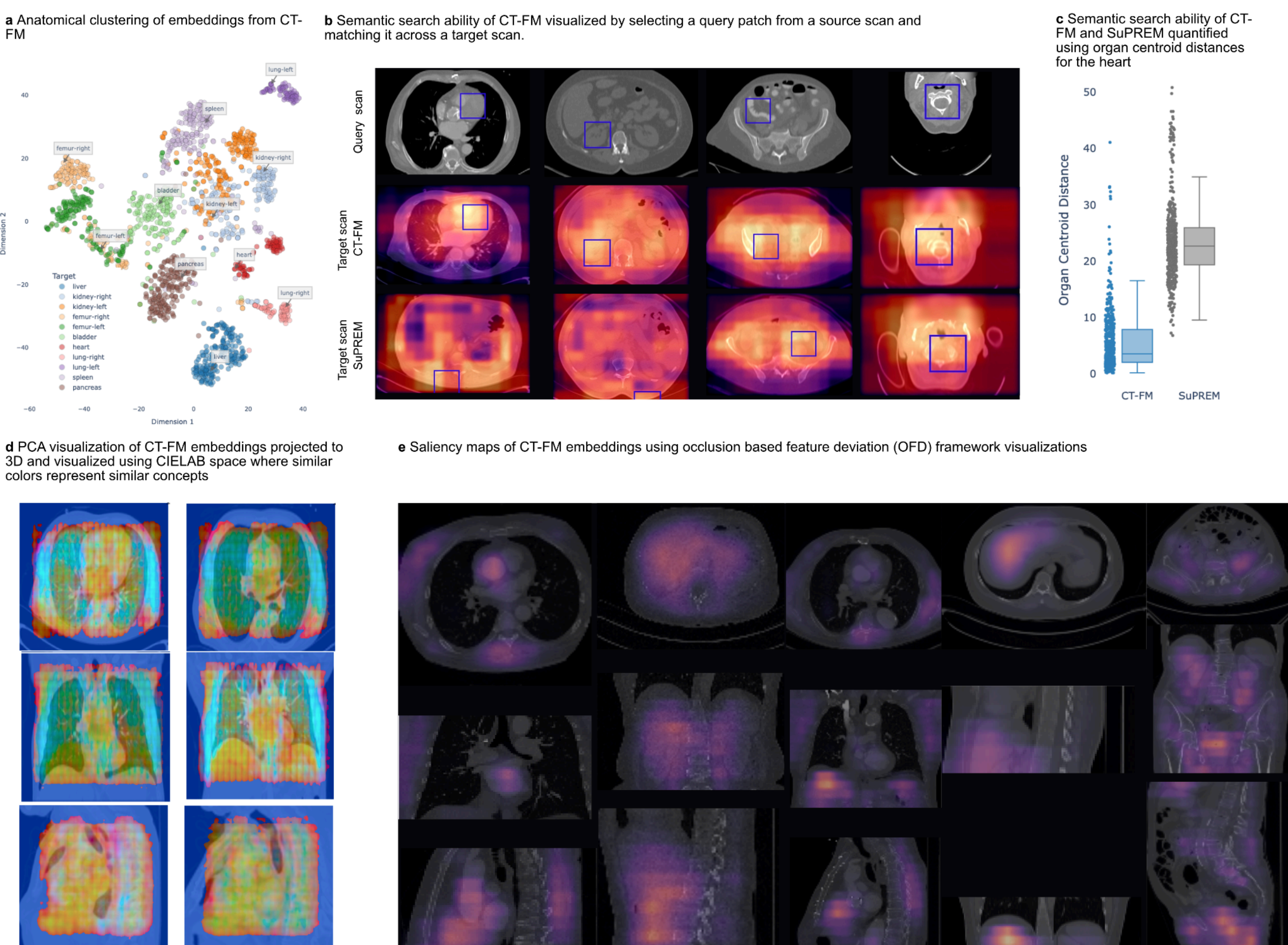

**Figure 6: Visualization and analysis of CT-FM embeddings.** (a) Anatomical clustering of embeddings, demonstrating the model's ability to group similar anatomical structures. (b) Semantic search example: a query patch from a source scan is used to find corresponding regions in target scans. CT-FM identifies more relevant regions than SUPREM. (c) Quantitative evaluation of semantic search using heart organ centroid distance. (d) PCA visualization of embeddings projected into 3D CIELAB color space, where similar colors represent semantically similar concepts. (e) Saliency maps using occlusion-based feature deviation (OFD), highlighting the regions in PET/CT images most influential to the CT-FM embeddings.

## PCA semantic analysis

To interpret high-dimensional embeddings of CT-FM, we performed dimensionality reduction and visualized the first three principal components by assigning each a color space[29]. Similar anatomical structures exhibited consistent colors across scans—for example, the heart appeared red, lungs blue, and bones green—illustrating the linkage between feature values and anatomical concepts (see **Figure 6d**).

## Saliency and stability of CT-FM features

Applying a saliency method designed for feature representations—relying on maximal deviations in the feature space due to occluded regions—we mapped features back to specific locations in the CT scans, identifying salient regions corresponding to key anatomical areas such as the lungs, heart, diaphragm, vertebral column, and pelvic girdle (see **Figure 6e**), which indicates that the model focuses on structures across the respiratory, cardiovascular, gastrointestinal, and musculoskeletal systems. Our CT-FM embeddings also  demonstrated robustness across test-retest CT scans in the RIDER cohort, showing high similarity in corresponding patches despite variations in acquisition parameters, and can detect outliers due to larger positioning differences (see **Extended Data Figure 7**).

### Pre-training ablations

Our pre-training methodology was validated through systematic ablation studies: introducing intra-sample objective modifications improved micro Dice scores in SimCLR, SimSiam, and VicReg by 0.257, 0.3246, and 0.154, respectively, and increasing the number of contrastive crops in SimCLR from N=5 to N=15 enhanced micro Dice scores from 0.654 to 0.732 (see **Extended Data Figure 6**). Our pre-training representations were also validated during training to explore the quality of representations learnt across various training checkpoints; training longer does not necessarily yield optimal representations for transfer, as our best macro- and micro-averaged Dice scores were achieved at epoch 449 (see **Extended Data Figure 6**).

## DISCUSSION

In this study, we introduced CT-FM, a large-scale 3D image-based foundation model with a pre-training design tailored for radiological interpretation tasks. Leveraging a dataset of 148,000 CT scans from the Imaging Data Commons and employing a task-agnostic self-supervised learning strategy, we demonstrated that CT-FM outperforms several state-of-the-art baselines across a range of relevant tasks. These tasks include whole-body anatomical segmentation, heterogeneous tumor segmentation, head CT triage classification, and micro- and macro-scale medical image retrieval. Additionally, we explored the interpretability and semantic richness of the embeddings learned by CT-FM, highlighting its inherent anatomical awareness and stability.

Foundation models - either generalist or specialist - rely strongly on learning representations using large-scale datasets across different modalities. In the context of medicine, the development of a General Medical Artificial Intelligence (GMAI) requires flexible interactions via models pre-trained independently and jointly on several modalities that are encountered in medical practice[30]. However, several challenges exist in the development of such models as medical images have unique characteristics whose features and patterns differ significantly from those in natural images[31] requiring considerations beyond those of generalist design. Recognizing these considerations, we develop a CT domain foundation model, CT-FM through a native 3D image-based contrastive pre-training promoting awareness of 3D structure. CT-FM demonstrated strong performance in fine-tuning and zero-shot embedding tasks providing a robust and adaptable 3D vision foundation model to further the development of medical imaging foundation models. Previous efforts in this domain, have predominantly focused on contrastive methods for 2D image-based/video-based pre-training[5,6,32,33] or joint image-text pre-training[5,6,32] as well as supervised methods for tasks requiring anatomical awareness[17,20]. Learning representations natively in 3D with structure-aware pre-training allows us to perform well on segmentation tasks, outperforming several baselines prominent in medical segmentation literature for whole-body and heterogeneous tumor segmentation, which have been challenging for generic contrastive pre-training methods[20]. Decoupling from textual representations, which are highly contextual and could contain large amounts of boilerplate text aids our model in capturing rich semantic representations solely from information inherent to volumetric medical data. During pre-training CT-FM encountered only a fixed pixel spacing and slice thickness but was able to adapt successfully across diverse spacings present in each of the segmentation datasets likely due to the nature of augmentations applied in contrastive pre-training. CT-FM also surpassed Merlin, when fine-tuned on the same data subset, another recent electronic health records and vision joint pre-trained foundational model, highlighting the effectiveness of our vision-centric pre-training strategy. With respect to classification, we adapted CT-FM to predict the triaging of head CTs where very few samples of such anatomy were seen in the pretraining dataset. Regardless, CT-FM provided a benefit over the random baseline across both in-distribution and out-of-distribution datasets, while it improved over SuPREM in out-of-distribution generalization. In the medical image retrieval task, CT-FM significantly improved compared to baselines. These retrieval results highlight CT-FM's capability to facilitate meaningful and precise similarity searches. This could support diagnostic decision-making by enabling efficient case retrieval and cohort identification in clinical workflows.

Beyond task-specific performance, interpretability is essential for the clinical adoption of AI models. Our analyses revealed that CT-FM's embeddings exhibit inherent anatomical clustering and semantic concept identification. The model demonstrated the ability to link specific anatomical structures across different scans, as evidenced by semantic concept searches and principal component analysis visualizations. These findings suggest that CT-FM learns to perform tasks while having developed a nuanced understanding of anatomical relationships within CT volumes, which is crucial for building clinical trust. Stability and robustness are critical in medical applications, where variability in patient positioning and scanning parameters is common. Our test-retest analysis on the RIDER dataset showed that CT-FM's embeddings are consistent across repeated scans, indicating reliability in varying acquisition settings and where inconsistent, the cause was registration inaccuracies enabling the spotting of outliers. This robustness suggests that CT-FM can be effectively integrated into real-world clinical workflows.

Despite these promising results, our study has limitations. The computational resources required for training and fine-tuning such a large model limited us from employing more extensive cross-validation or ensemble methods, which could have further improved performance in downstream tasks. Our pre-training strategy in CT-FM involved several ablations against generic pre-training as well as ablations for design choices within the new strategy. However, as the design space is quite large, we were only able to explore an intuitively guided subset of the space. A wide range of comparisons and tasks were incorporated to validate the efficacy of CT-FM but quantification abilities such as diagnostic and prognostic remain to be explored[10]. For qualitative comparisons, we selected scans randomly and demonstrated the performance of our model as practical considerations limited us from exploring all relevant data qualitatively.

Future work could focus on developing a comprehensive medical imaging foundation model by integrating CT-FM with textual representations, leveraging limited supervised data, and incorporating additional modalities. This holistic model could be applied to clinically relevant use-cases, including assistive report generation, radiotherapy structure segmentation and planning, visual question answering, and large-scale medical image retrieval. While the present pre-training strategy is not confined to CT scans, extending its applicability to other imaging modalities such as MRI, PET, X-rays, and ultrasound scans would enhance its utility in the medical imaging domain. The integration of CT-FM with diverse semantic and imaging modalities could be achieved through a multi-stage pre-training paradigm, utilizing CT-FM as the foundational visual encoder and incorporating unstructured radiology reports, structured electronic health records, and supplementary clinical data. Exploring hybrid models that combine supervised and self-supervised learning strengths could yield even more robust representations.

In conclusion, CT-FM represents a significant advancement in developing foundation models for medical imaging. As a pre-trained model, CT-FM enables more efficient use of available data, improved performance in low-data regimes, and robust generalization across heterogeneous datasets. In settings where annotated data is scarce, CT-FM can be fine-tuned with minimal additional labels to achieve robust performance. As an embedding framework, it facilitates search and retrieval functions and outlier detection, aiding in diagnostic decision-making and research applications. Overall, our study highlights the potential of self-supervised learning in enhancing radiological interpretation by pre-training on a large-scale 3D CT dataset and demonstrating superior performance across diverse tasks. In a commitment to advancing research and ultimately improving patient care, we are open-sourcing our data, code, and model weights. This will enable the community to build upon our work and drive innovation in this critical field.

## METHODS

**Study Population.** We utilize several datasets across this study for both pre-training and downstream evaluation.

Our pre-training dataset is sourced from the Imaging Data Commons data repository and contains 148,000 CT scans from 81148 studies and 32643 patients. Our dataset is chosen through quality-based selection criteria and includes 69 different cohorts with various inclusion criteria (see Supplementary Information) A majority of the dataset is comprised of the National Lung Cancer Screening Trial (NLST) cohort and the detailed composition can be found in Supplementary Table 3. .

TotalSegmentator[21] dataset comprises scans collected from 1368 CT examinations of the University Hospital Basel Picture Archiving and Communications System from 2012, 2016, and 2020. The CT series were sampled randomly from these examinations, resulting in a dataset of 1228 CT scans. The scans were annotated for 117 structures using the Nora Imaging Platform for manual segmentation by physicians. To speed up the annotations, models for existing structures and iterative learning were leveraged. Each examination was finally reviewed at the end and corrected when necessary. The dataset was split into 928 training, 52 validation, and 248 testing scans. We used the default spacing provided in the dataset of 1.5mm cubic. We used a different split for comparison against the Merlin CT foundation model, where the same split as the compared study was chosen.

Medical Segmentation Decathlon[23] is a challenge dataset acquired across multiple institutions, anatomies, and modalities for real-world clinical applications. We use four tumor cohorts: the hepatic, hepatic tumor adjacent to vessels, pancreas, and lung as part of the datasets. The hepatic dataset comprises 201 contrast-enhanced CT images from patients with primary and metastatic liver cancer from IRCAD Hopitaux Universitaires, Strasbourg, France, where liver tumors were annotated. The hepatic vessel dataset consists of 443 portal-venous phase CT scans of patients with primary and metastatic liver tumors from MSKCC with heterogenous tumors adjacent to vessels being annotated. The pancreas dataset has 420 portal-venous phase CT scans of patients with pancreatic mass resections from Memorial Sloan Kettering Cancer Center (MSKCC), New York, USA, where pancreatic masses were annotated. The Lung dataset contains pre-operative scans for 96 patients with NSCLC from the TCIA, where tumors within the lung are annotated. We used custom splits for each task as the test set is made private for the challenge. The preprocessing parameters for each dataset were determined dynamically through the Auto3DSeg pipeline.

SinoCT[34] is a dataset collected from Stanford University that comprises 9,779 head CT scans with a balanced distribution of normal (45%) and abnormal (55%) cases, as classified through expert radiological interpretation. We adopted the best-performing pre-processing approach from the SinoCT authors, stacking four CT images with distinct windowing settings (window level - WL, window width - WW): blood (WL 40, WW 80), subdural (WL 25, WW 300), stroke (WL 32, WW 8), and bone (WL 600, WW 3000).[25] Rather than using channel-wise stacking, we adopted width-wise concatenation to maintain compatibility with pre-trained networks like CT-FM and SuPREM, which expect single-channel grayscale inputs. Images were standardized through resampling to 2x2x5mm$^3$ voxel spacing and were center-cropped to a patch size of 128x128x48, with zero padding applied when necessary. This approach guaranteed complete head coverage in a single patch provided to the network, eliminating the risk of missing critical pathologies at patch boundaries while maintaining computational efficiency and ensuring consistent spatial context across all samples. We split the dataset into training, validation, and test sets using a ratio of 8:1:1, resulting in 7,823 samples for training, 978 for validation, and 978 for testing.

CQ500[26] dataset comprises 491 head CT scans collected from multiple radiology centers in New Delhi, India. Initially, three radiologists annotated the scans across various pathology categories, including intracranial hemorrhage, fracture, midline shift, and mass effect. These annotations were subsequently relabeled into a binary normal/abnormal classification. The dataset exhibits a balanced distribution, with approximately 45% normal and 55% abnormal scans, similar to the SinoCT dataset. We applied the same preprocessing techniques used for SinoCT to the CQ500 dataset. All 491 scans were utilized exclusively for testing purposes, ensuring an independent evaluation of the model's performance on this external dataset.

OrganMNIST3D[27], a subset of the MedMNIST dataset, originally sourced from the Liver Tumor Segmentation Benchmark Dataset (LITS'17), is used in this study. The dataset consists of 1743 samples of organs present in abdominal CT scans. 3D bounding boxes from 11 organs[35] are used to crop 3D volumes containing organ FOVs and formulated as a multi-class classification problem. 131 scans in the training set are also used, and 70 CT scans from the source test set are treated as the test set. Since we use this dataset for retrieval, we only use a train and test set.

3D-MIR[28] dataset derived from the Medical Segmentation Decathlon (MSD), includes diverse lesion types. Each lesion is meticulously segmented and described, emphasizing both its morphological characteristics and its anatomical context within the associated organ, making it suitable for quantitative analysis, retrieval, and other advanced medical image analysis tasks.

**Pre-training of the CT Foundation Model.**

The pre-training of the CT Foundation Model used a modification of the original SimCLR framework. The SimCLR framework involves two key concepts - 1) Learning invariance to augmentations/transformations of the sample, and 2) Differentiating views of the sample from "negative" views, primarily to avoid representational collapse. In the original framework, negative views for a sample are chosen as all elements in the batch other than the view itself. In **Extended Data Figure 8a**, we show a case in natural-world images where elements in a batch contain similar elements.

Consider $B = \{x_1, x_2, \ldots, x_n\}$, where $B$ is a batch of size $n$ sampled from the data distribution $x_i \sim D, \quad \text{for } i = 1, 2, \ldots, n.$. Next, transformation functions $T_A$ and $T_B$ are applied to instances $x_n$ to generate augmented views, where transformations are sampled from a collection of transformations as $T_i \sim \mathcal{T}, \quad \text{for } i = 1, 2, \ldots, n.$

$$x_1^A = T_A(x_1); \quad x_1^B = T_B(x_1)$$

The views are passed through $f_\theta$, typically an encoder structure with weights $\theta$ to obtain a high-dimensional embedding representation $z$

$$z_n = f_\theta(x_n)$$

In our example setting, shown in **Extended Data Figure 8a**, the similarity between embeddings between views of the same instance, $z_1^A$ and $z_1^B$ is approximately equal to the similarity between views across separate instances indexed by $n$.

$$\text{Sim}(z_1^A, z_1^B) \approx \text{Sim}(z_1^A, z_n^A) \quad \forall n \neq 1$$

Such behavior occurs when covariances between views over the dataset are of the same order as the variances across the instances of the dataset.

$$\text{Cov}(x^A, x^B) \approx \text{Var}(x)$$

This likely leads to suboptimal representation quality. Notably, in 3D medical imaging use cases, variations across batch elements might not differ vastly from variations across sample views, mimicking the example setting. To address this, we propose a modification where the negative views for a sample are composed of patches from within the sample itself. This modification is inspired by previous works, such as PatchNCE[36], in which patches are contrasted against others from the same instance. In contrast to PatchNCE, which is used in a generative framework, our implementation is used for variance maximization as an auxiliary objective in popular contrastive learning frameworks. Our implementation can be formulated by considering $x$ to be a 3D grid of elements,

$$x = \{p_{i,j,k} \mid i = 0, \ldots, I;\ j = 0, \ldots, J;\ k = 0, \ldots, K\}$$

where $p_{i,j,k}$ denotes an element in the 3D space of $x$, and $I$, $J$, and $K$ define the bounds of $x$ along each dimension. Each patch $h$ is a sub-volume of $x$ with a fixed size of $(s_i, s_j, s_k)$. A patch centered at position $(i, j, k)$ can be defined as:

$$h = \{p_{i\prime,j\prime,k\prime} \mid i\prime \in [i, i + s_i - 1],\ j\prime \in [j, j + s_j - 1],\ k\prime \in [k, k + s_k - 1]\}$$

Let $\mathcal{T}(x)$ represent the distribution of all patches in $x$. Then, a patch $h$ is sampled from $x$ as $h \sim \mathcal{T}(x)$. If we sample multiple patches independently from $x$, we can represent a set $S$ of sampled patches as:

$$S = \{h_m\}_{m=1}^{M} \quad \text{where} \quad h_m \overset{\text{i.i.d.}}{\sim} \mathcal{T}(x)$$

where $M$ is the total number of patches sampled. $S$ is used instead of a mini-batch where we first create two augmented views $\tilde{h}_{m,1}$ *and* $\tilde{h}_{m,2}$ using random data augmentations. Let $z_{m,1} = f_\theta(\tilde{h}_{m,1})$ *and* $z_{m,2} = f_\theta(\tilde{h}_{m,2})$ be the representations of the two augmentations of patch $h_m$, where $f_\theta$ is the encoder with a projection head. For the positive pair $(z_{m,1}, z_{m,2})$ from patch $h_m$, the SimCLR contrastive objective is formulated as,

$$\ell_{m,1,2} = -\log \frac{\exp(\text{sim}(z_{m,1}, z_{m,2})/\tau)}{\sum_{k=1}^{2M} \mathbf{1}_{[k \neq m]} \exp(\text{sim}(z_{m,1}, z_k)/\tau)}$$

The total contrastive loss over all patches in $S$ is given by:

$$\mathcal{L} = \frac{1}{2M} \sum_{m=1}^{M} (\ell_{m,1,2} + \ell_{m,2,1})$$

In addition to modifying the objective, we make several decisions in our framework, such as generating views with a relatively small patch size to choose a certain concept level to learn invariance across. The positive views are formed for this concept level, and negatives are chosen across the scan to differentiate among concepts at that level. We term this Intra-sample contrastive learning and implement it across various contrastive pre-training methods. First, several patches are randomly sampled from within a CT scan with a chosen patch size of 24x128x128 (z,y,x) in image dimensions. As preprocessing, scans were resampled to 3mm slice thickness and 1mm in-plane resolution. The selection of the patch size was made considering the actual physical dimensions of anatomy in the image, with a cube of size 12.8cm x 12.8cm x 6.4cm chosen as the concept level to learn invariance across roughly the length, width and thickness of the human heart. We choose a patch from the randomly sampled patches and transform it into two augmented views with augmentations selected from standard contrastive learning choices (such as random resize and crop) and medical imaging-specific choices (such as histogram shifting, random affine transformations, and intensity scale variations.)

A SegResNet[37] encoder obtains latent representations from the inputs following the generation of views. The SegResNet architecture was chosen due to its strong performance across several medical imaging segmentation tasks, as well as due to its asymmetric encoder-decoder structure, which is preferred for representation learning. The latent representations formed using fully convolutional blocks have a dimensionality of 512, after which we use a projection head to map these representations to the manifold space used for learning similarity/dissimilarity in the SimCLR objective. The default SimCLR objective, a normalized temperature-scaled cross-entropy loss, is used with cosine similarity to quantify the similarity between representations. Similarity between representations of views from the same patch is maximized, while similarity between views of one vs the rest is minimized.  To prioritize decreasing similarity between the most dominant differences, a temperature value of 0.1 is selected (default value). **Extended Data Figure 8b** shows a visual diagram of our pre-training pipeline.

Our model was pre-trained for 500 epochs using 20 patches from a single scan and patches across 16 different scans for 320 patches in a single batch. The learning objective was applied independently across 20 patches from each scan but averaged over the collection of scans. This ensured that representations learned within a scan were not biased to specificities present in the single scan. The training was performed on 4xNVIDIA Quadro RTX 8000 GPUs with distributed data parallel splitting the data uniformly across the GPUs. We used the Adam optimizer with a learning rate of 0.0001 and weight decay of $10^{-6}$ with a warmup cosine scheduler using 10 warmup epochs. The training took 24 days owing to the large amount of training samples. During training, we selected the preferred checkpoint, epoch 449, by evaluating the quality of representations learned for few-shot semantic segmentation on the TotalSegmentator dataset. A lightweight decoder was trained on top of the learned representations, and performance was compared on a held-out set of samples. **Extended Data Figure 6** shows a comparison of different epochs and their performances.

**Ablation of pre-training strategies:** We implemented pre-training strategies from three broadly categorized families: Deep Metric Learning (DML), Self-distillation, and Correlation-based Methods. From DML, we chose the SimCLR framework; from self-distillation, we chose SimSiam, and from correlation, VicReg. For each of these strategies, we applied the intra-sample modification described previously. Since SimSiam and VicReg do not use negatives, the modification involved composing the batch using patches from a single scan. We also fill the batches with patches from other scans but apply the loss criterion only over elements of the batch that originate from the same scan. The loss criterion is then averaged across all scans that compose the batch. SimSiam's criterion effectively remains the same but with a different batch composition process. Each framework was trained for 25 epochs following default parameters indicated in the original papers. Transformations, optimizers, and learning rate scheduling were kept consistent across frameworks to reduce the influence of variability in representation learning. At the end of 25 epochs, we evaluated the performance of each of the frameworks on a small subset of the TotalSegmentator dataset by freezing the learned encoder and adding a decoder on top to predict segmentation masks. While we explored several different pre-training frameworks, we eventually chose SimCLR due to its ability to explicitly model dissimilarity across elements through the selection of positives and negatives.

**Adaptation of the Vision Foundation model**

The CT vision foundation model was adapted for each segmentation and classification task by fine-tuning end-to-end (i.e., adapting all layers). For the segmentation tasks, we added the decoder of the SegResNet back as the pre-training only trained the encoder. For classification, we added an adaptive max pooling layer, fully connected layers with ReLU activations, and an output layer. Specific configurations were chosen for the task as appropriate and are described below.

**Whole body segmentation adaptation:** We used a custom fine-tuning algorithm designed with choices presented in the TotalSegmentator paper and in MONAI's Auto3DSeg framework. The algorithm was implemented using the Lighter configuration framework[38] built on MONAI[39] and Pytorch Lightning. A baseline was also trained using the same algorithm. Our fine-tuning framework was trained for 300 epochs using the AdamW optimizer with a learning rate of 0.0002 and a patch size of [96, 160, 160]. The criterion optimized was a weighted combination of Dice score and cross-entropy loss. An effective batch size of 8 across 4xRTX8000 GPUs was chosen, with a batch size of 2 on each GPU. A warmup cosine learning rate scheduler was used during the optimization. Each sample in a batch was sampled from the uniform distribution of all 117 labels. During training, a series of augmentations - affine transformations, gaussian noise, smoothing, intensity scaling, and shifting. For limited data training, we randomly selected 5, 10, 20, 50, and 100 volumes from the training set.

The validation set was also reweighted according to the training percentage. As we use the same data split as described in the VISTA3D[17] paper, we add comparisons against Auto3DSeg, VISTA3D and nnUNet from the results presented.

**Multi-region tumor segmentation:** We chose the Auto3DSeg framework for adapting our foundation model to tumor segmentation as it has demonstrated competitive performance across numerous challenges and tasks. Its design philosophy aims to provide a comprehensive and versatile pipeline for developing automated segmentation methods. Given the inherent complexities of tumor segmentation, such as heterogeneity, data imbalance, and relatively small sizes, Auto3DSeg remains the optimal choice of framework. Therefore, we refrained from developing our solution and used a well-established framework instead, as the focus was on evaluating the efficacy of our model pre-trained weights. Our model was plugged directly into the Auto3DSeg framework and finetuned with a reduced learning rate to allow better persistence of pre-trained weight spaces[17]. Our compared baseline was trained using all default parameters of the Auto3DSeg framework with the same model architecture as our pre-trained model. This allowed us to compare the benefits offered by our pre-trained model fairly. Specifically, we chose a learning rate of 0.0005 for our fine-tuning while baselines used the default Auto3DSeg learning rate of 0.0002.

**Head abnormality binary classification:** The head CT triage framework was implemented using the Lighter framework integrated with MONAI. The model architecture consisted of two main components: a feature extraction backbone and a classification head. The backbone was either SegResNet for CT-FM and random initialization models or UNetEncoder for SuPREM. The classification head processed the backbone's feature maps through a cascade of operations. First, a 3D adaptive average pooling layer reduced spatial dimensions to 1×1×1, followed by flattening and a two-stage fully connected network. The first dense layer reduced the embedding dimensionality by half and incorporated ReLU activation, while the final linear layer outputs a single binary logit. CT volumes underwent preprocessing with four distinct intensity windowing operations optimized for different anatomical features: blood (0 to 80 HU), subdural (-125 to 175 HU), stroke (28 to 36 HU), and bone (-900 to 2100 HU). Following intensity windowing and concatenating them width-wise, we applied random affine augmentations during training with a probability of 0.2, incorporating rotations of ±0.26 radians and scale variations of ±0.2 along all three axes. The model was optimized using Binary Cross-Entropy with Logits Loss and the AdamW optimizer, configured with an initial learning rate of $1\times10^{-4}$ and weight decay of $1\times10^{-5}$. A cosine learning rate schedule was implemented to gradually decay the learning rate over 40 epochs. The training utilized distributed data parallelism across four GPUs with a per-GPU batch size of 16. Model checkpoints were saved based on validation AUROC performance and evaluated at the end of each epoch.

**Content-based Retrieval of organs and lesion characteristics:** For the retrieval tasks, we followed a search and retrieve procedure of generating embeddings a priori for the entire training dataset. Test embeddings were computed on the fly and compared with the training embeddings. For the OrganMNIST3D, a single embedding was computed by passing the entire resampled FOV (default size provided in the dataset) to CT-FM. For 3D-MIR, patch-wise embeddings were first computed and then aggregated to form a single embedding per scan - similar to the baseline proposed by the 3D-MIR authors[28]. We found that across several aggregation strategies, taking the minimum value of embeddings across all patches provided the best retrieval. The difference across different retrieval methods was, however, not very large. The search used a cosine similarity metric between the test embedding and the pre-computed corpus of train embeddings. Top-k matches with the highest cosine similarity were retrieved and used for subsequent analysis.

**Analysis Metrics**

Our algorithms were evaluated using metrics relevant to the use cases chosen. Multi-class segmentation of the entire body was evaluated using macro-averaged Dice score across all the labels and specific label groups. Individual label Dice were also compared and presented across models. For tumor segmentation, we used the Dice of the predicted tumor segmentation and average surface distance, a stringent metric that evaluates the efficacy of the predicted segmentation boundary. Regarding head abnormality classification, a binary classification problem, we use AUC and F1 scores to evaluate our methods. Finally, we use average precision, hit rate, and F1 score to evaluate retrieval methods.

**Semantic Concept Clustering**

Semantic concept clustering relies on clustered low-dimensional projections of our high-dimensional embeddings. We use t-stochastic Neighbourhood Embeddings with different perplexity settings to project our 512-dimensional embedding to 2 dimensions. In the lower dimensional space, we color the embedding (point) corresponding to the label it represents. As our clustering dataset contains 11 anatomical organs with contexts such as "located in the abdomen," "left or right," "located nearby," etc, we can meta-analyze the clusters formed by embeddings from several methods and determine the best approach.

**Semantic Concept Search**

An extension of the semantic concept clustering and content-based retrieval of instances leads us to explore semantic concept search methods where content-based retrieval is performed at a micro-scale on semantic concepts present within an instance, such as different organs located in a scan. To enable this search, we implement a framework (see **Extended Data Figure 9**), available on our Github repo, where an end-user can select a 3D patch in a source scan of any arbitrary FOV and execute a semantic search on one or several target scans. The framework first computes the embedding for the selected patch. It compares it with embeddings generated from all possible patches in the target scans using a sliding window mechanism to generate such patches. The compared patches with the least cosine distance or highest cosine similarity are chosen as the best match patch, generating a heat map that shows cosine similarity rankings for compared patches. The end-user can then scroll through the target scans to explore regions of high similarity across the entire FOV. The evaluation was performed by  selecting scans with the same body part DICOM tag from the TotalSegmentator dataset.

**Organ Centroid Distance**

To quantitatively assess our semantic concept search, we introduced a metric called the Organ Centroid Distance (OCD). The OCD measures how accurately our method can locate a specific organ based on a semantic match. Essentially, we start with a known point within an organ (e.g., the heart) on a source scan. Then, our semantic search finds the best matching region in a target scan. The OCD is the distance between the true center of the target organ and the center of that best-matching region found by our algorithm. We computed the OCD across the entire TotalSegmentator dataset by selecting the heart as the organ from a randomly selected scan that imaged the thoracic region and contained a 3D segmentation for the heart and compared against all other scans in the dataset that met the same criteria.  Intuitively, the smaller the OCD, the better our semantic search is at pinpointing the correct anatomical location from a solely a single reference point.

### PCA analysis of learned semantic concepts

We selected two scans with a similar FOV covering the chest and extracted embeddings for patches of size 24x64x64 from each of the two scans. Each patch embedding is considered a sample, and a principal component analysis is performed on all samples to reduce the dimensionality of the aggregated embeddings from 512 to 3. The first principal component was thresholded to remove the background as it captures much of the variance. Following the removal of the background, we mapped values of patches within the foreground to the CIELAB color map, with the first principal component representing lightness and the second and third representing a and b components of red-green and blue-yellow. The CIELAB colormap was chosen due to its properties as a perceptually uniform space where a change in human perceived color represents a similar change in underlying numerical values. We then overlaid the CIELAB color map for principal components on top of the original image and compared it across the chosen scans.

### Salient regions captured by the embeddings

Several saliency methods have been implemented for class predictors (single outputs). However, we would like to probe salient regions that influence the feature embedding in our CT-FM. Although methods such as RELAX have been proposed, we observed that their generated saliency maps are quite biased to selected hyperparameters; therefore, we propose an Occlusion-based Feature Deviation (OFD) saliency determination method. Here, we generate features by occluding a single patch of size 8x8x8 in an image of any arbitrary size. The occluding is done through a sliding window and features are computed for every occluded configuration. Finally, we compute the cosine distance between the original and occluded images. Increased cosine distance indicates an increased saliency of that region, contributing to the overall embedding.

### LLM Usage Disclosure

We disclose the usage of LLMs solely for editing and refinement of the content of the paper. All content was originally written by authors listed on the study. LLM generated content has been thoroughly verified and reviewed.

## ACKNOWLEDGEMENTS

The authors acknowledge financial support from NIH (H.J.W.L.A: NIH-USA U24CA194354, NIH-USA U01CA190234, NIH-USA U01CA209414, NIH-USA R35CA22052, and NIH-USA U54CA274516-01A1) and the European Union - European Research Council (H.J.W.L.A: 866504). This work also used GPUs provided by Jetstream 2 through allocation CIS240307 from the Advanced Cyberinfrastructure Coordination Ecosystem: Services & Support (ACCESS)[40] program, which is supported by U.S. National Science Foundation grants #2138259, #2138286, #2138307, #2137603, and #2138296.

## AUTHOR CONTRIBUTIONS

Study conceptualization: I.H., S.P., H.A.; Data acquisition, analysis, and interpretation: I.H., S.P., D.B., A.F.; Methodological design and implementation: I.H., S.P.; Statistical analyses: S.P.; Code and reproducibility: I.H., S.P.; Technical support and feedback: K.B.; Writing of the manuscript: S.P., I.H.; Critical revision of the manuscript: All authors; Study supervision: H.A.

## DATA AVAILABILITY STATEMENT

All datasets used in this study for both pre-training and evaluation are publicly available, ensuring full reproducibility of our results. The pre-training data were entirely sourced from the Imaging Data Commons (IDC)[41], with exact query parameters and versioning information provided in the supplementary material to enable identical data retrieval. For evaluation, we utilized the following datasets, which can be accessed through their respective publications: TotalSegmentator[21], Medical Segmentation Decathlon[23], SinoCT[34], CQ500[26], 3D-MIR[28], and OrganMNIST3D[27]. Researchers are encouraged to access these datasets directly to replicate or extend our work.

## CODE AVAILABILITY STATEMENT

All our code is publicly available on Github which can be accessed through our accompanying website [https://aim.hms.harvard.edu/ct-fm](https://aim.hms.harvard.edu/ct-fm). The repository includes pipelines for data preprocessing, pre-training of CT-FM, transfer learning across all demonstrated use cases, and evaluation scripts to generate comparison metrics. The training framework is implemented using our in-house open-source framework, Lighter, which is also available at [https://github.com/project-lighter/lighter](https://github.com/project-lighter/lighter).

## COMPETING INTERESTS

No competing interests to declare

# EXTENDED DATA FIGURES

**a** Top 25 improvements and decrements across structures when comparing CT-FM with random initialization

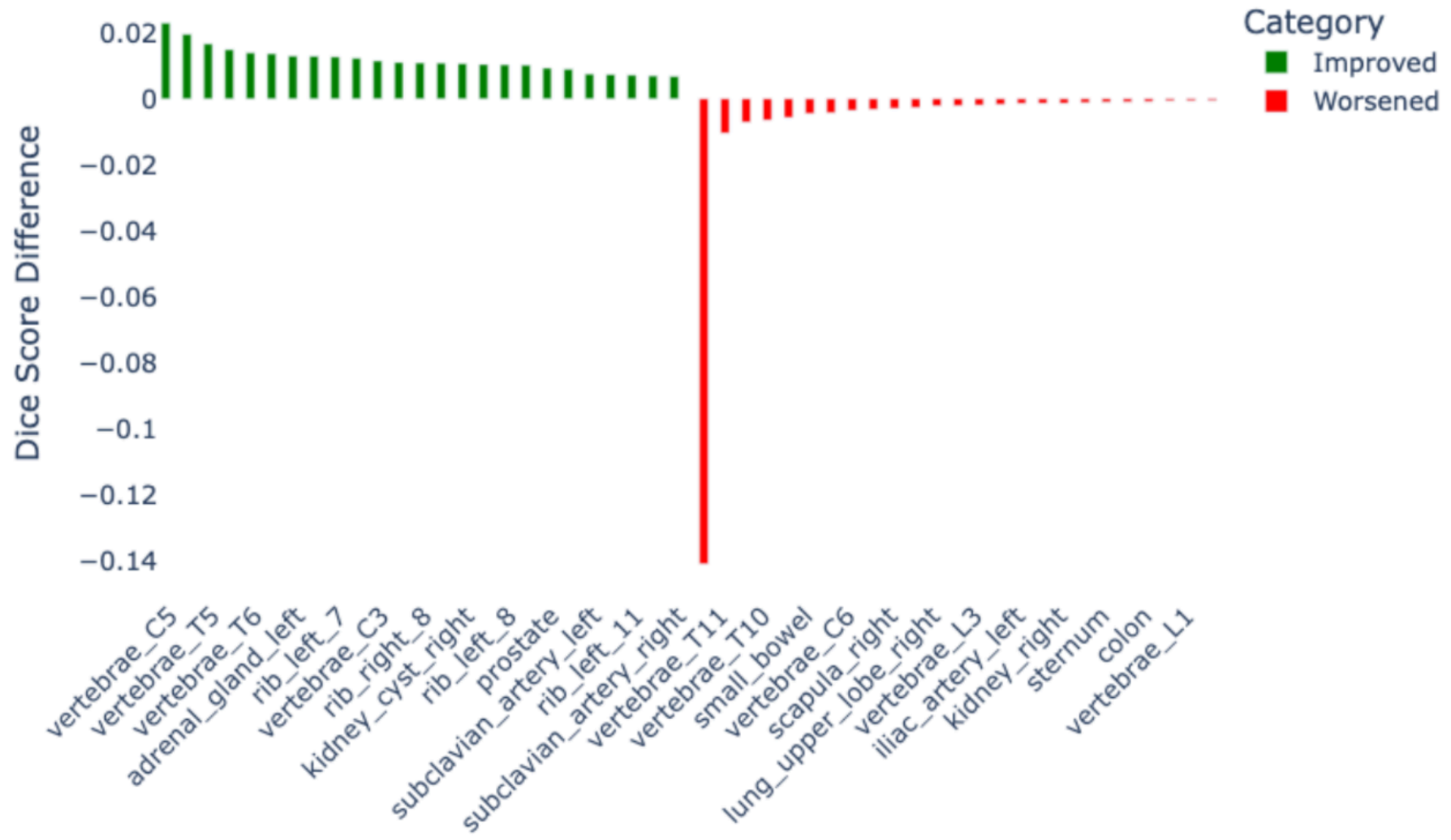

**b** Performance of models when comparing select structures with properties- large, smooth, small, irregular shape

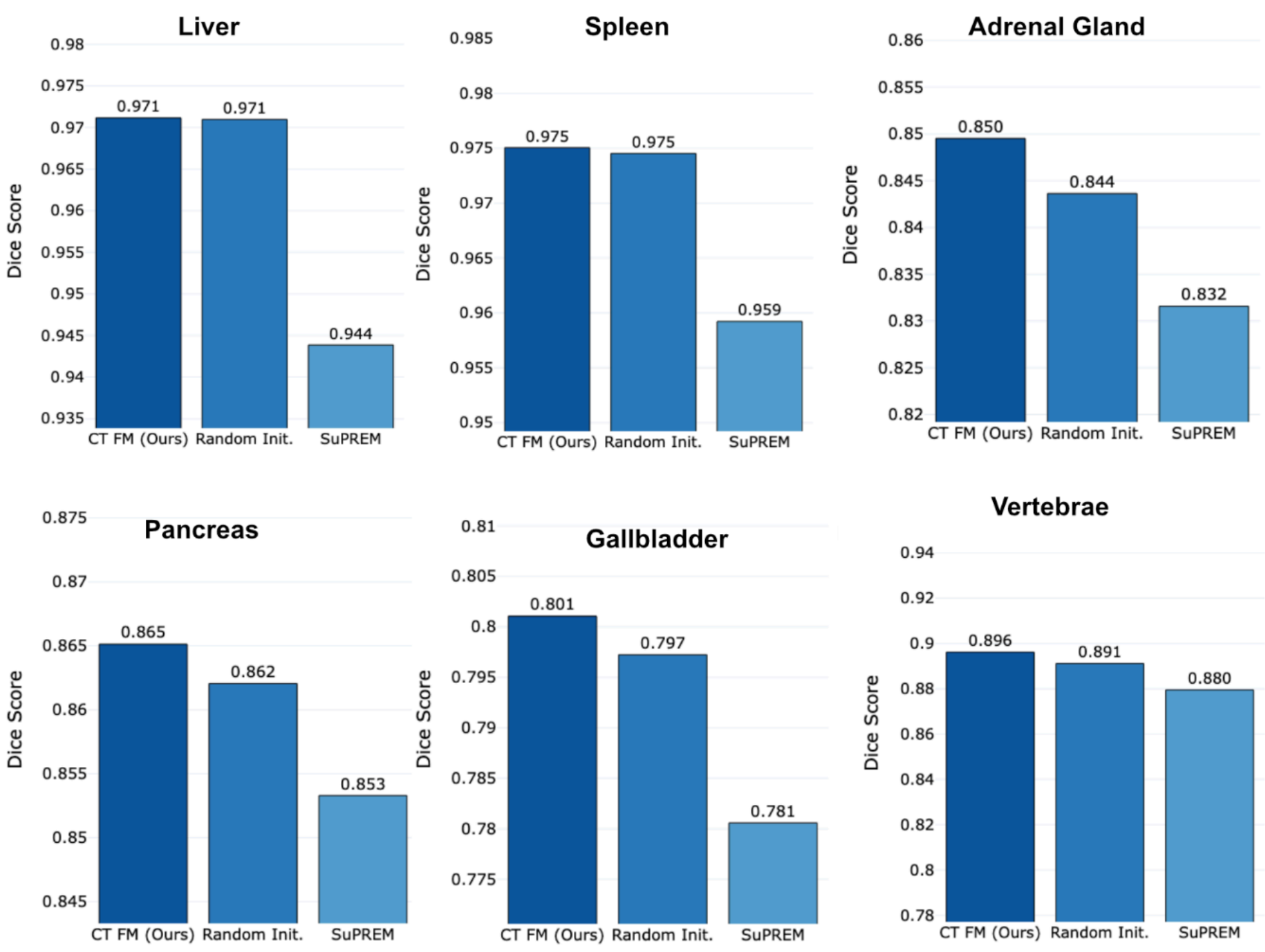

**Extended Data Figure 1:** Additional comparisons between CT-FM and baselines on segmentation of anatomical structures in whole-body CT scans.

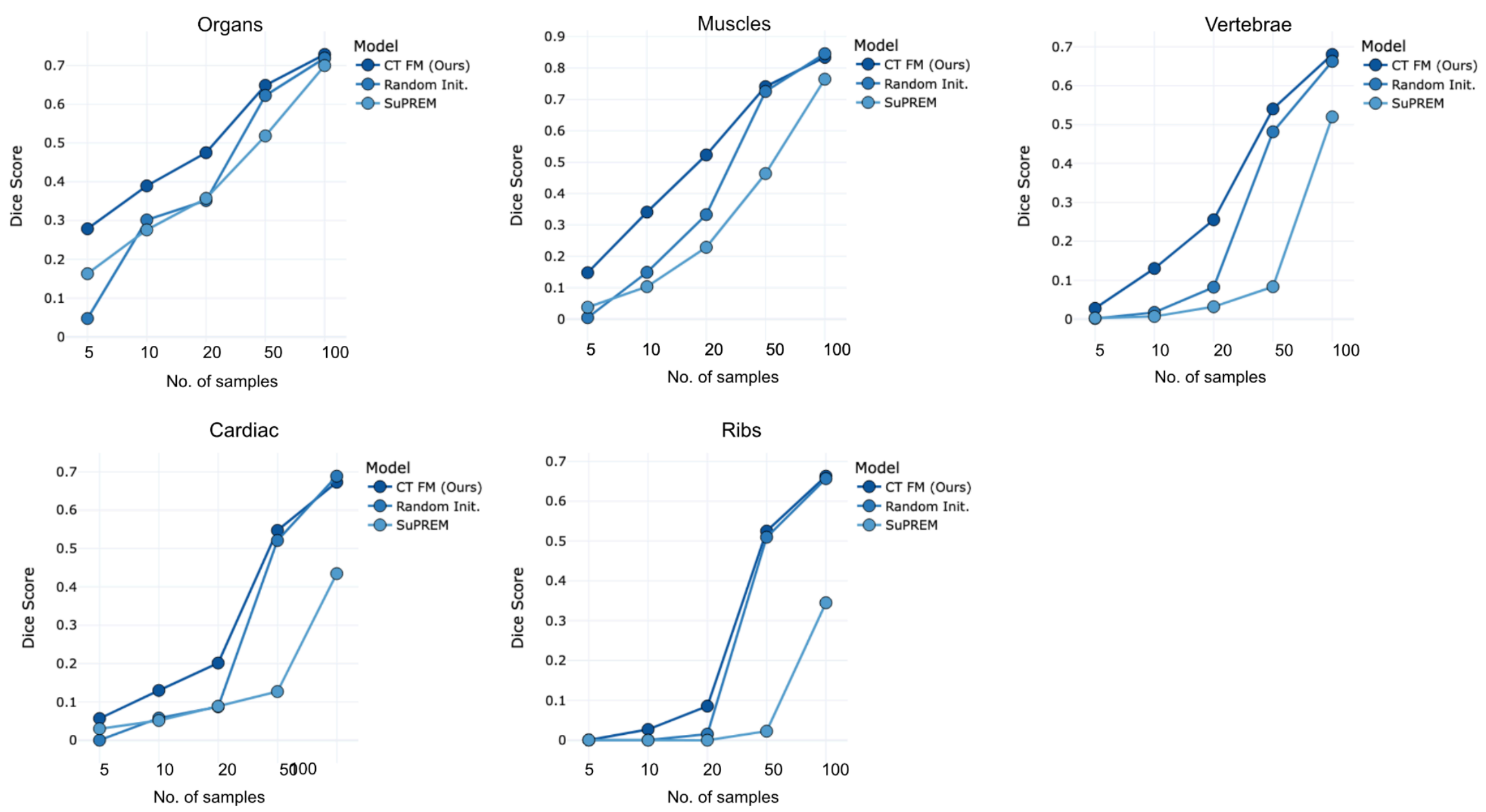

**Extended Data Figure 2:** Few-shot performance considering structure groups across the TotalSegmentator dataset

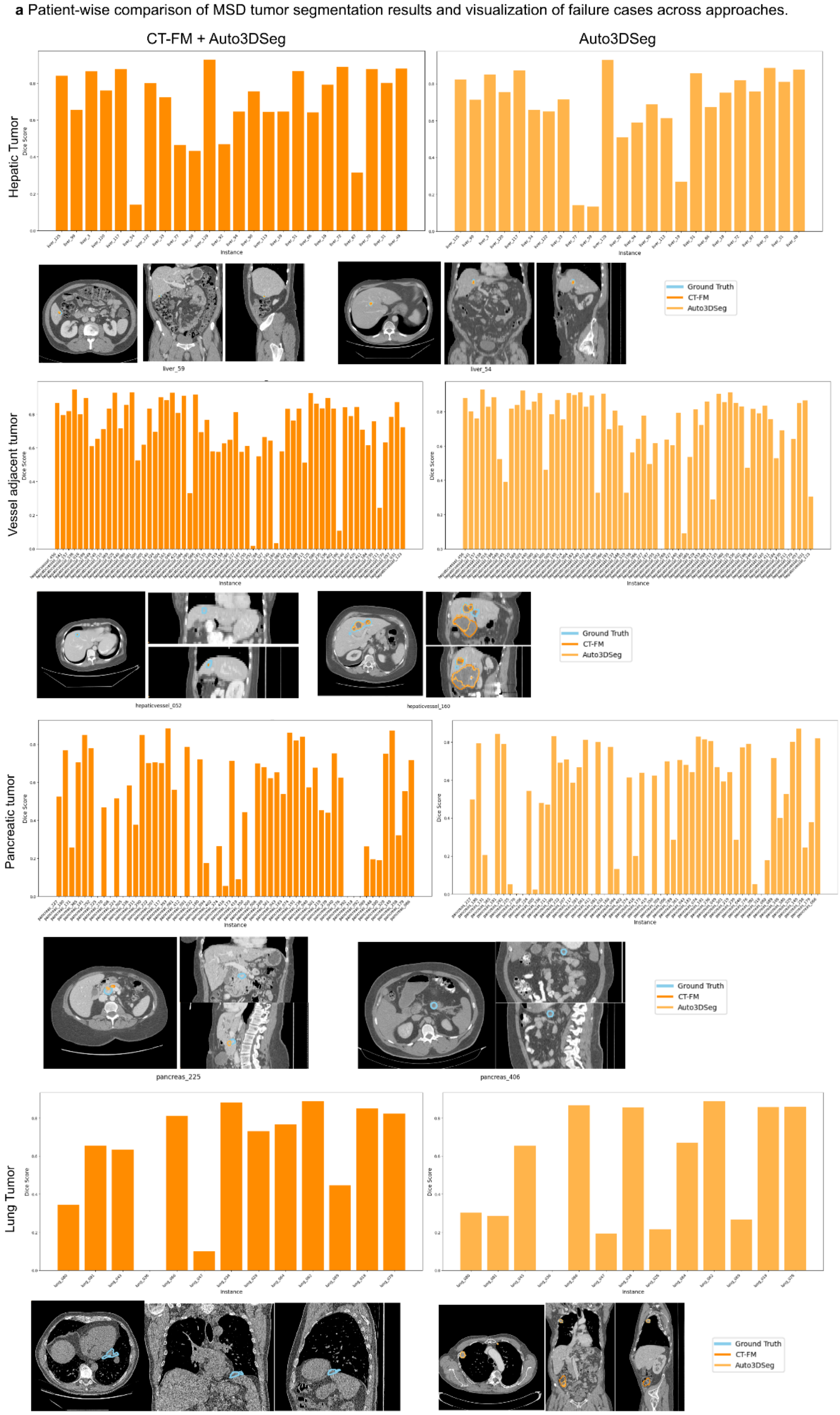

**Extended Data Figure 3:** Detailed patient-wise breakdown of comparisons on the MSD dataset with failure cases visualized.

**a** Content-based retrieval precision for k=3, 5 and 10 and overall average precision for the presence of a lesion on different anatomies of the 3D-MIR dataset

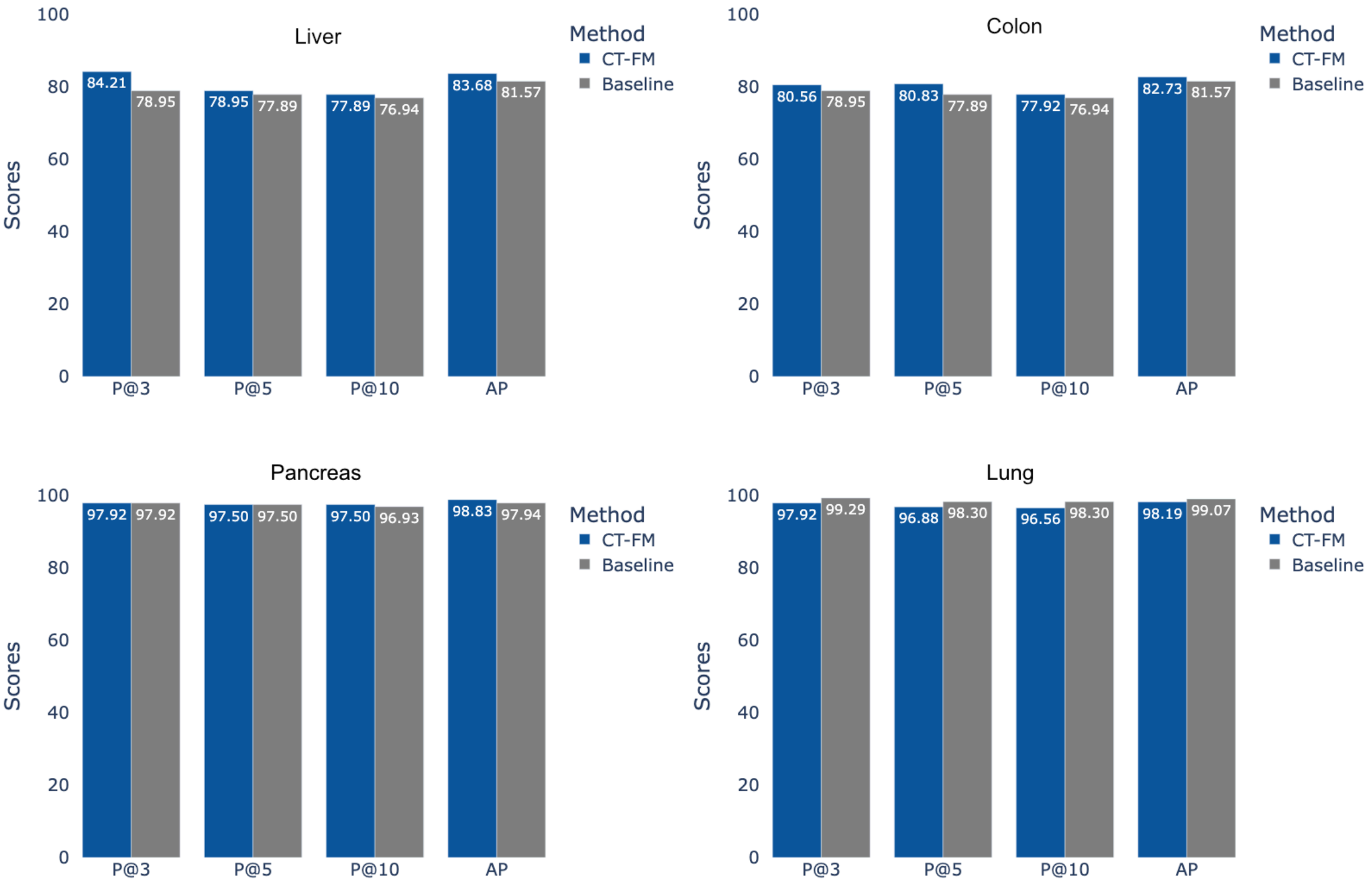

**Extended Data Figure 4:** Content-based retrieval detailed results for anatomical regions

**a** Comparison of anatomical clustering of CT-FM with baseline SuPREM across different values of perplexity

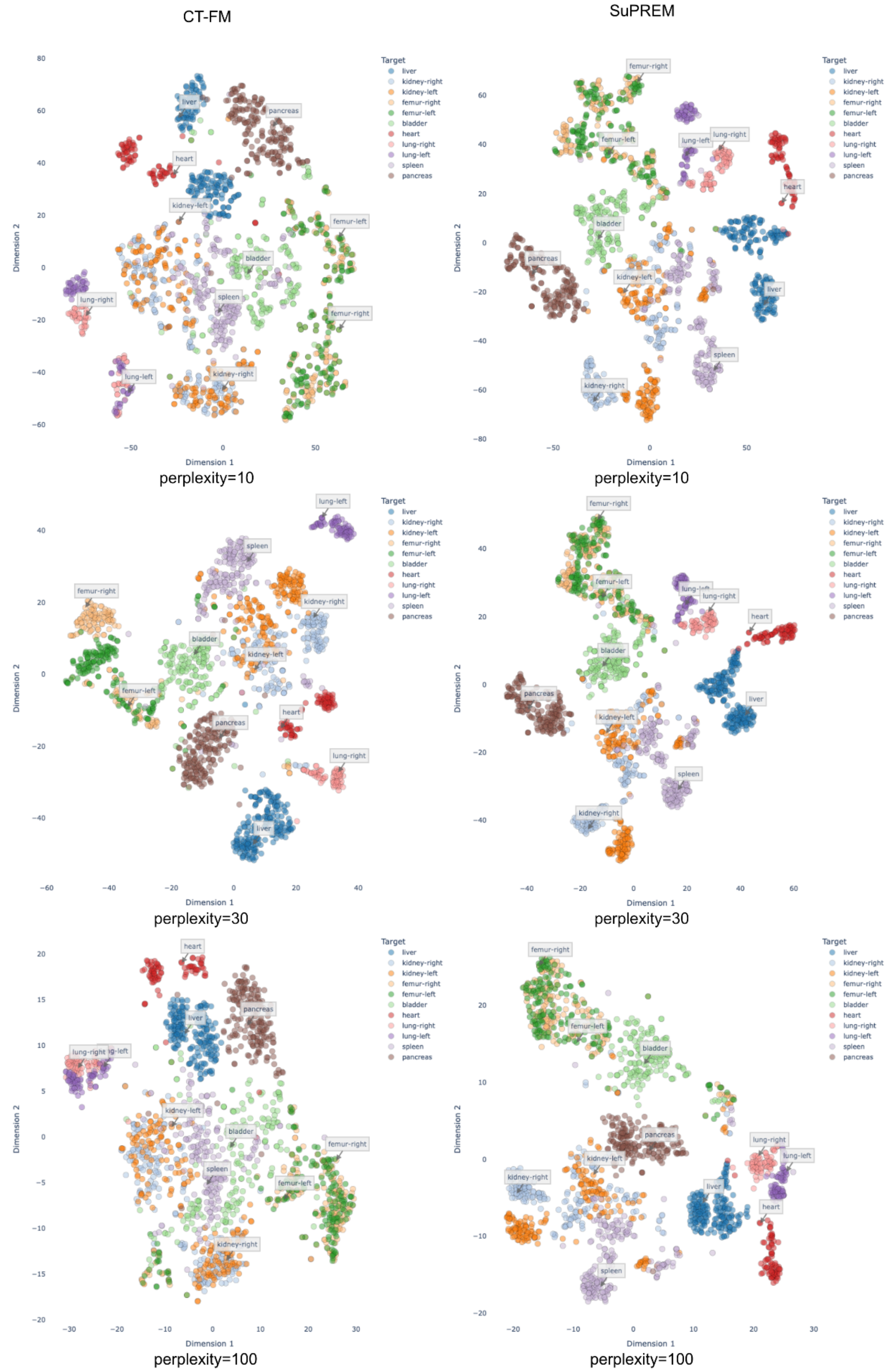

**Extended Data Figure 5:** Comparison of anatomical clustering between CT-FM and baseline

**a** Ablation of pre-training strategies evaluated on few-shot whole body segmentation on the TotalSegmentator dataset

**b** Ablation of number of crops on few-shot whole body segmentation

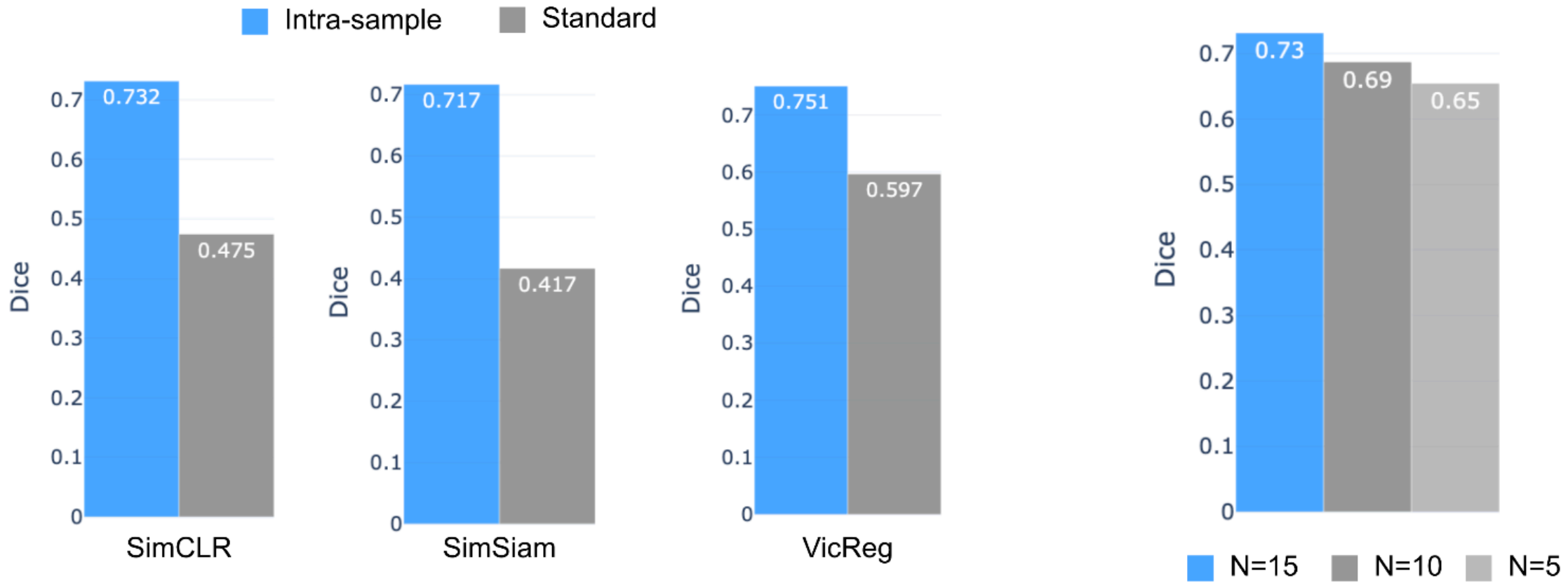

**c** Evolution of representation quality for whole body segmentation during the pre-training process.

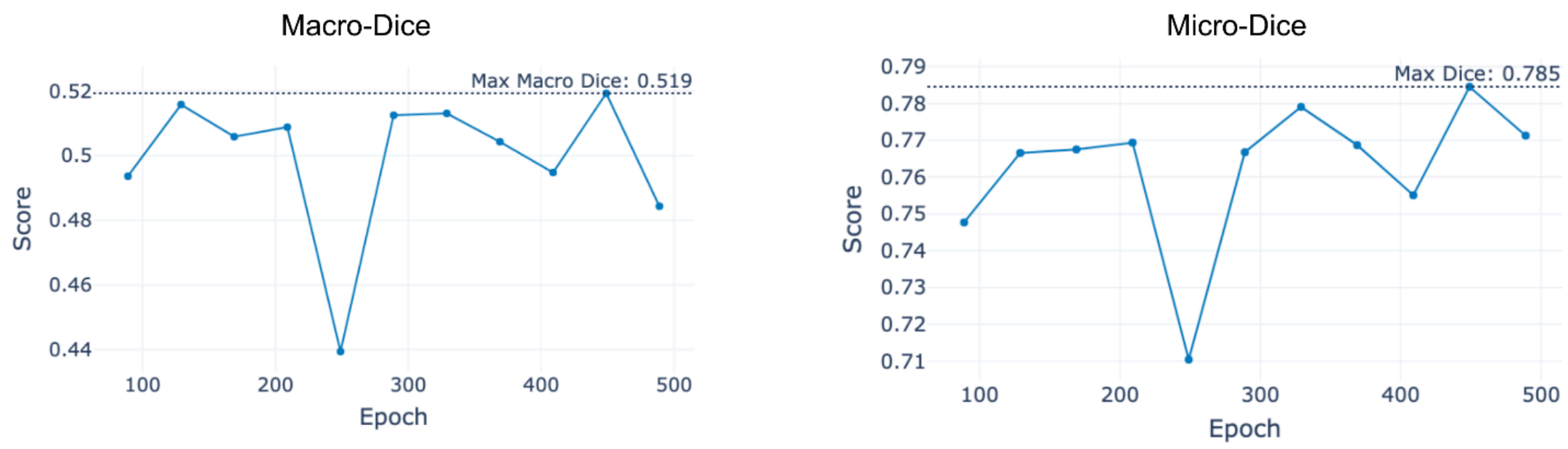

**d** Semantic concept matching between query patch and two selected target scans

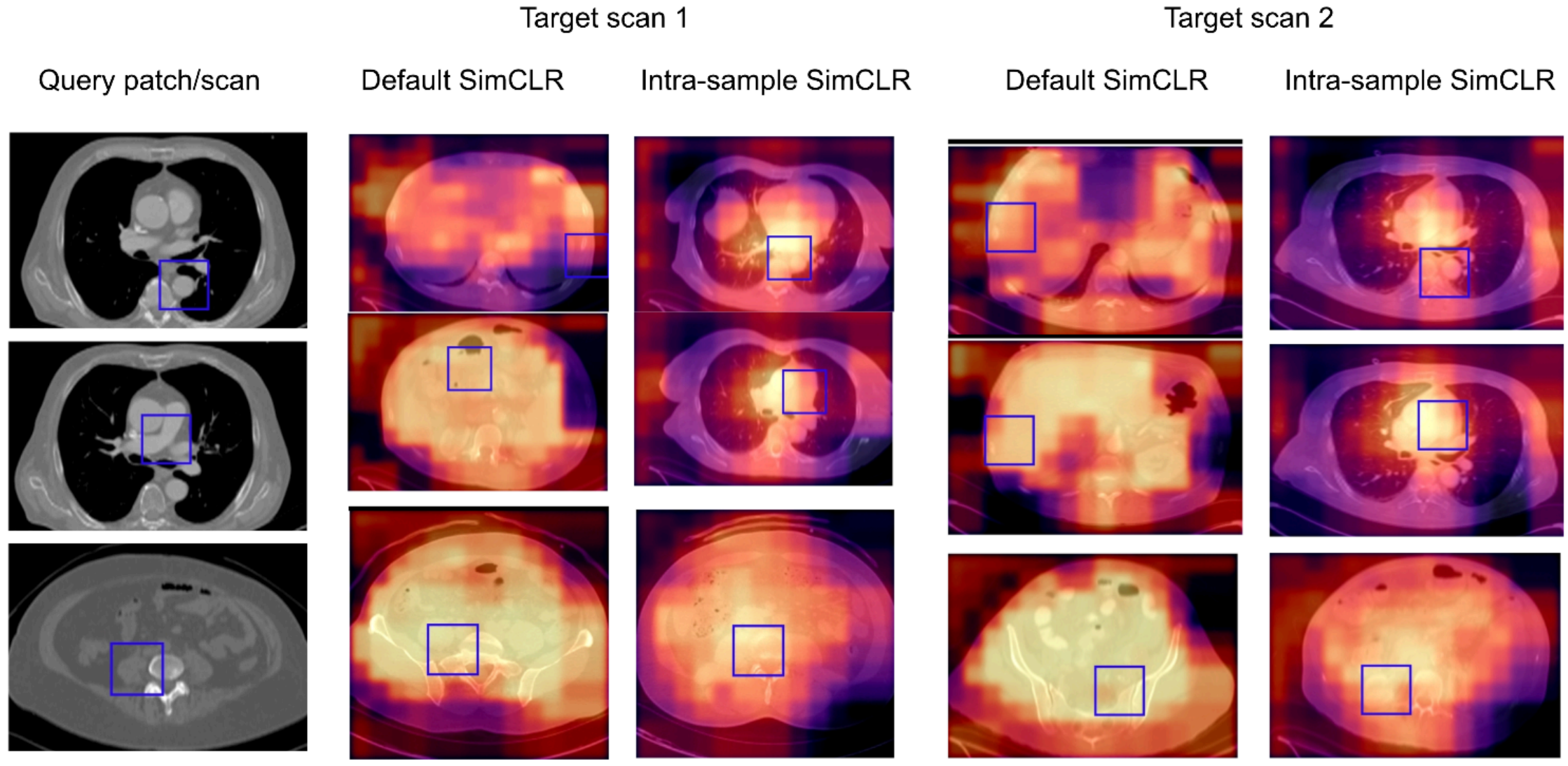

**Extended Data Figure 6:** Pre-training ablations and evolution of representation quality during pre-training

**a** Stability of patch embeddings on the RIDER test-retest dataset evaluated through cosine similarity

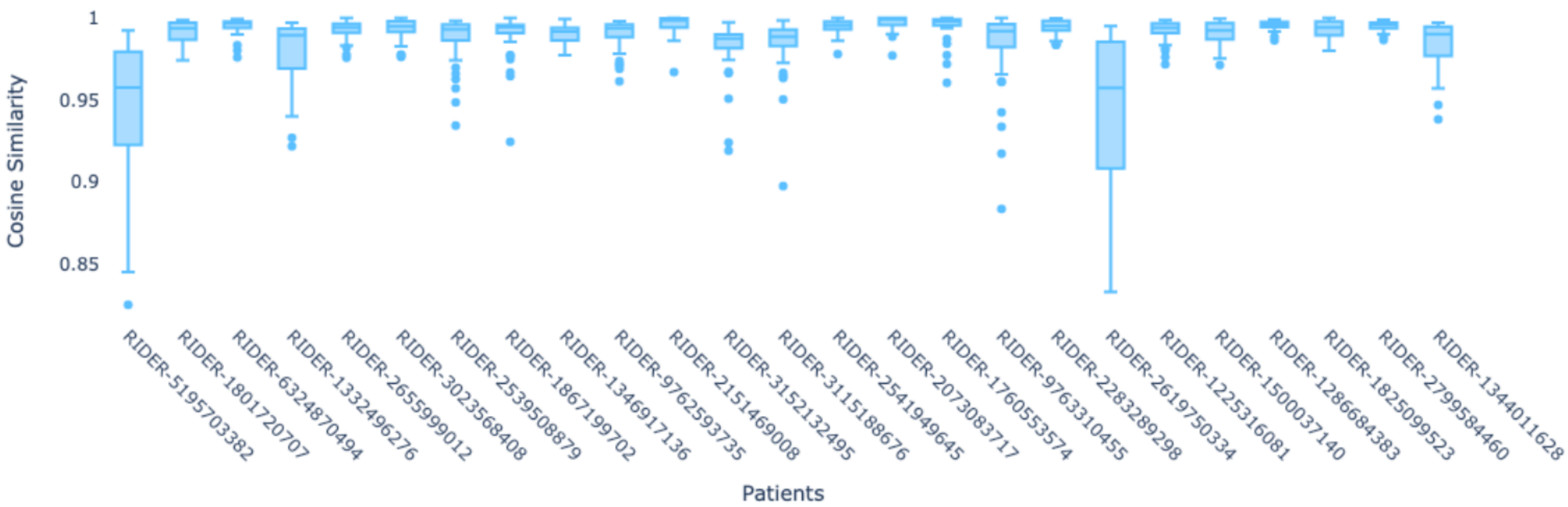

**b** Stability of patch embeddings on the RIDER test-retest dataset evaluated through mean squared error

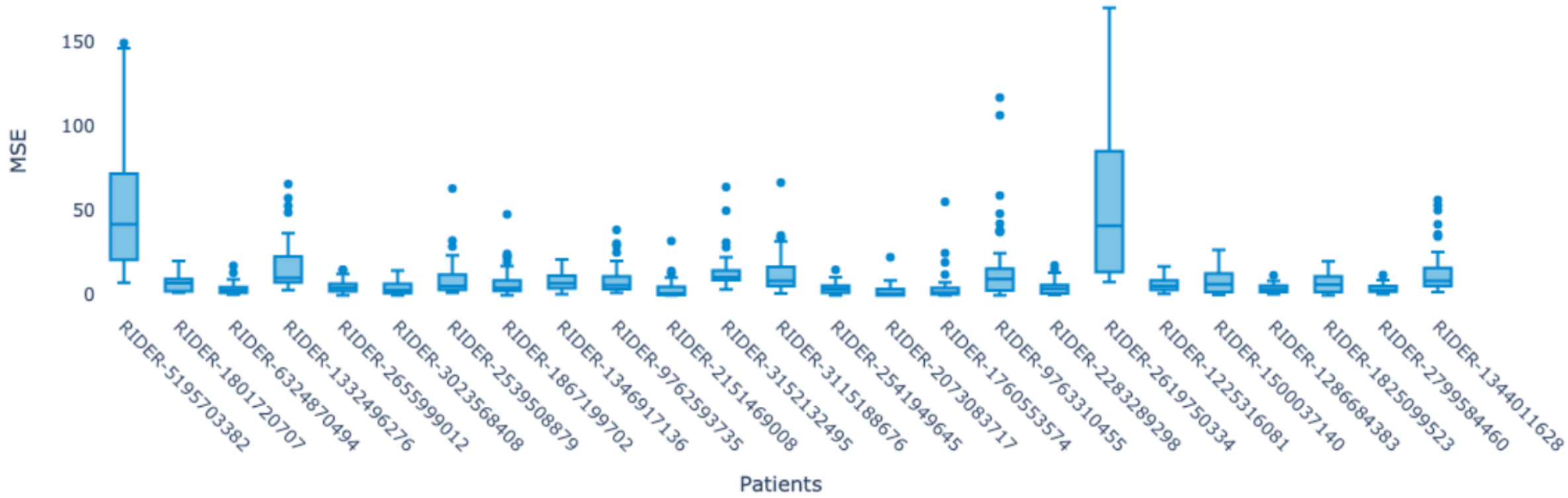

**Extended Data Figure 7:** Stability of CT-FM embeddings compared patch-wise between the RIDER dataset test-retest scans.

**a** Contrastive learning scenario for batch with limited variance between instances when compared to variance between views

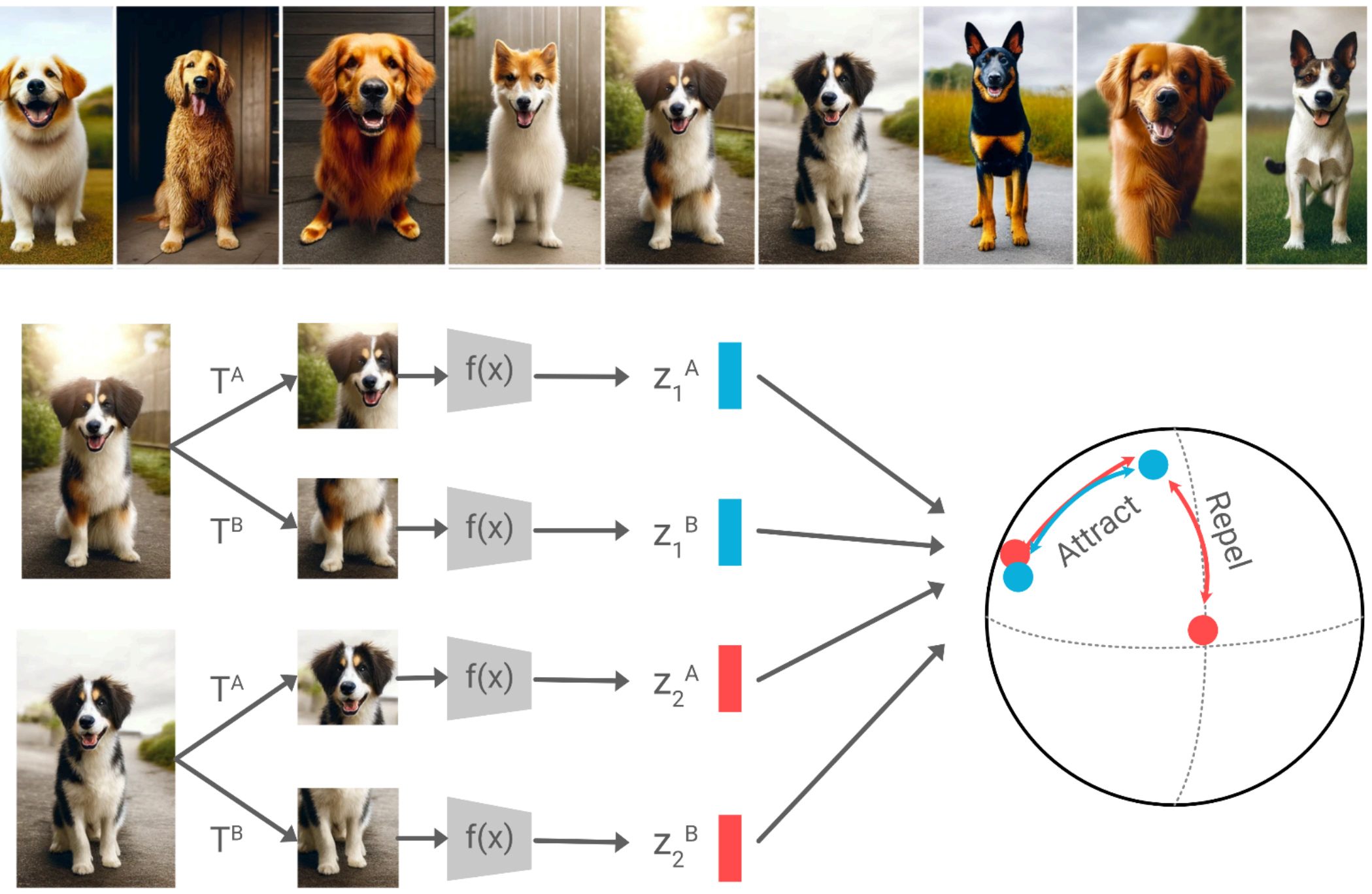

**b** Reformulation of the contrastive learning objective for limited variance medical imaging contexts

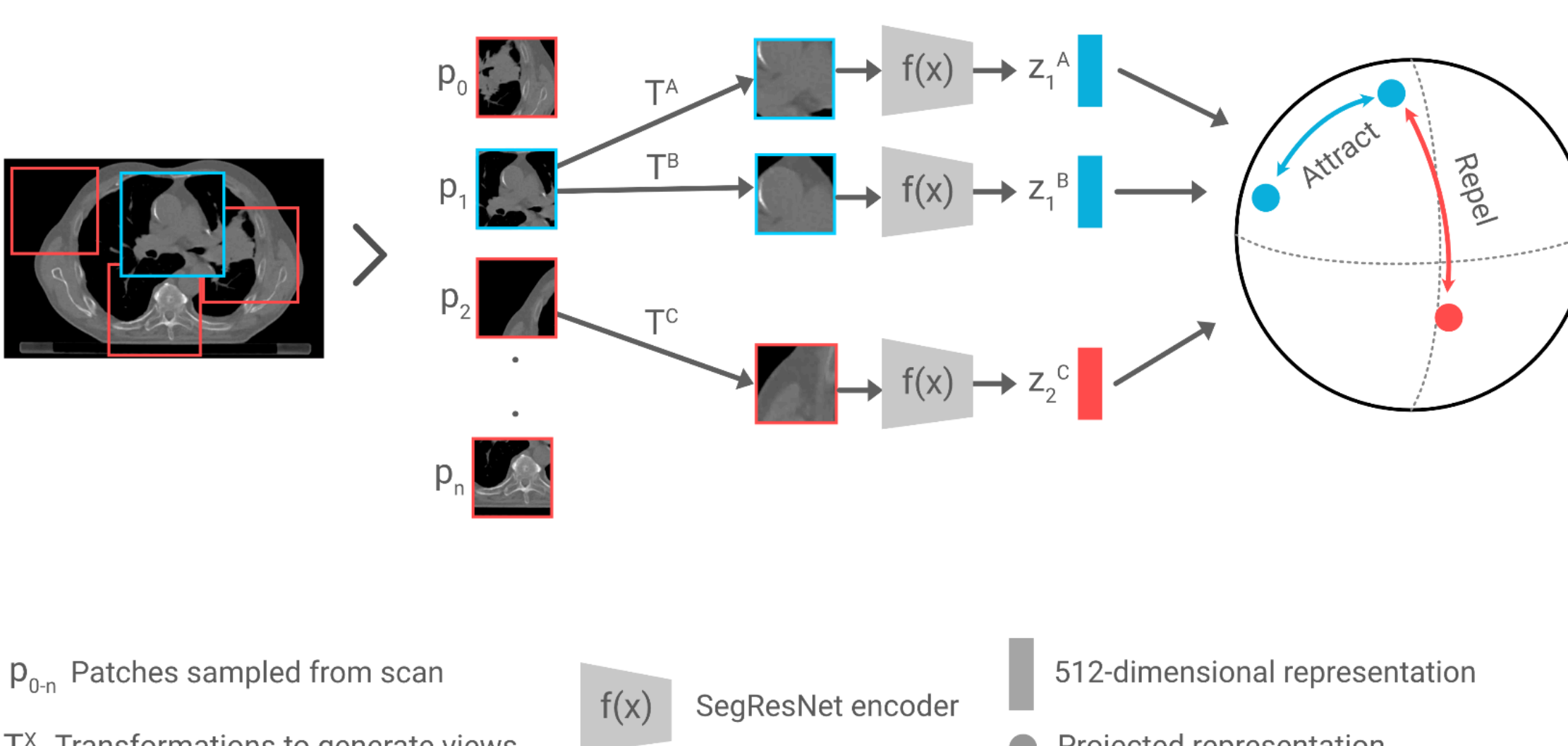

**Extended Data Figure 8:** Intra-sample pre-training adaptation for SimCLR

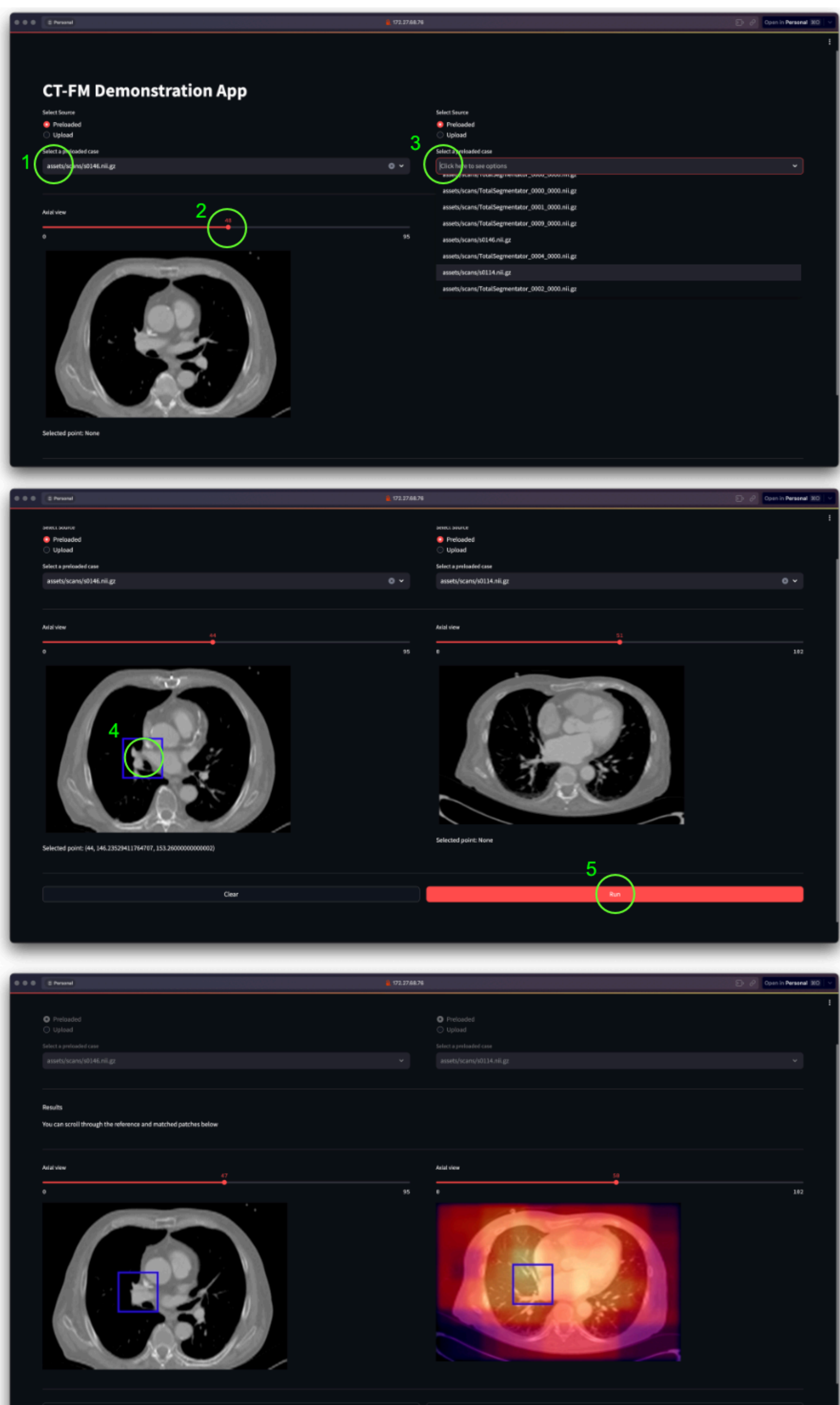

**Extended Data Figure 9:** Semantic search app workflow. Steps in the process are 1) select the source scan; 2) select the slice of interest; 3) select the target scan; 4) select a point around a region of interest in the source scan around which a 3D box will be cropped; 5) Run the semantic search framework to obtain the results

**a** Top 25 improvements and decrements across structures when comparing CT-FM with random initialization

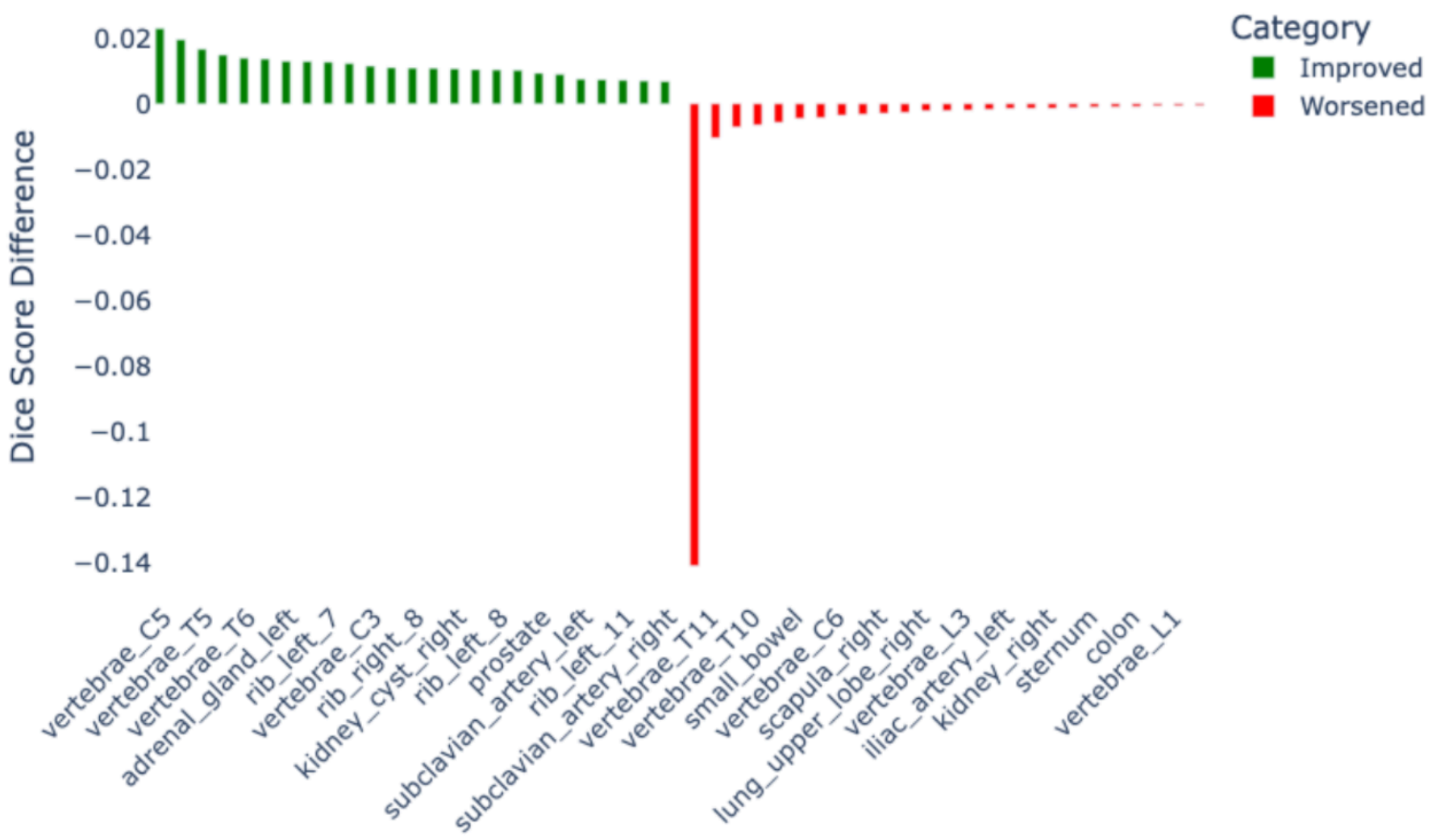

**b** Performance of models when comparing select structures with properties- large, smooth, small, irregular shape

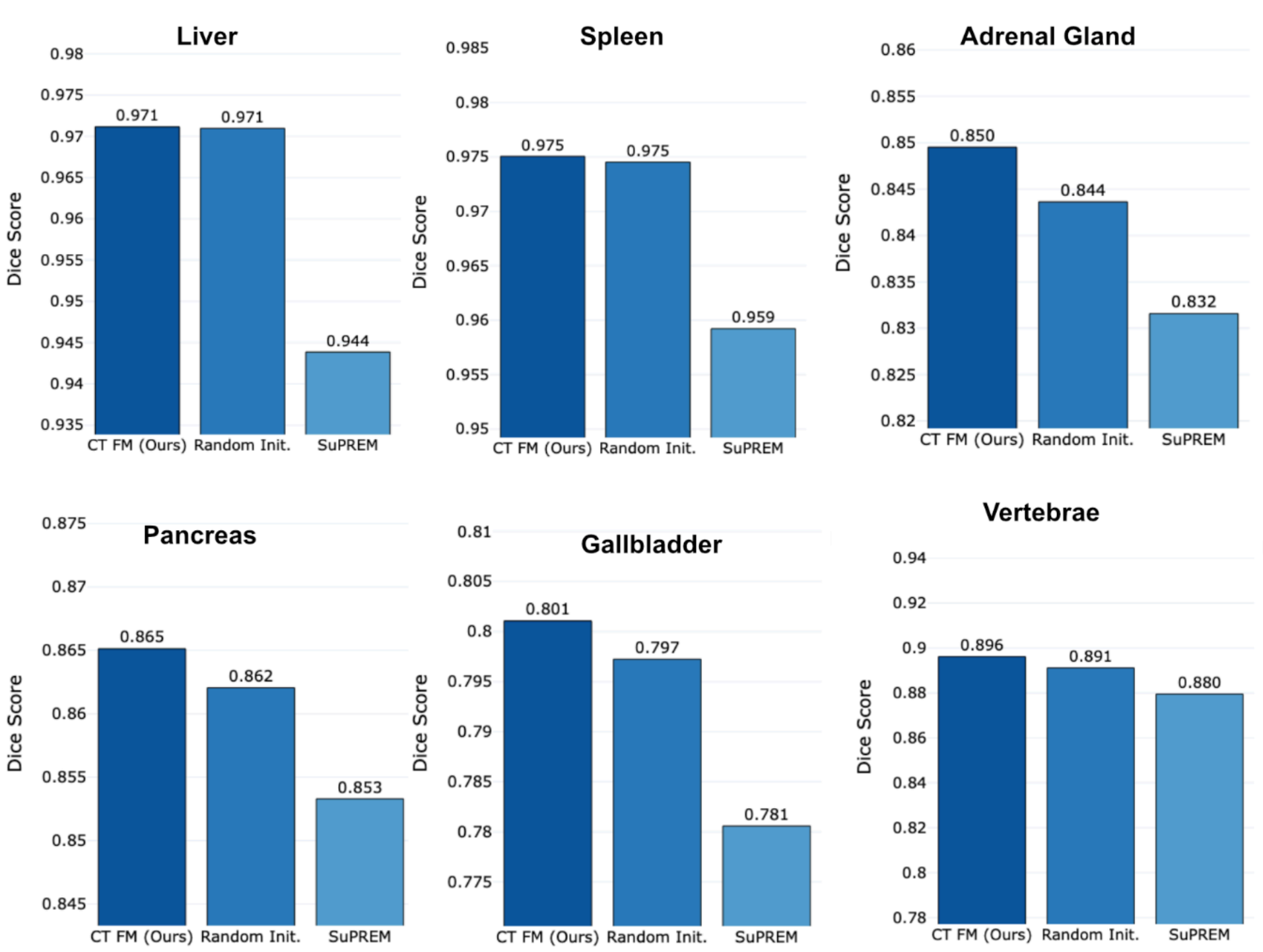

**Extended Data Figure 1:** Additional comparisons between CT-FM and baselines on segmentation of anatomical structures in whole-body CT scans.

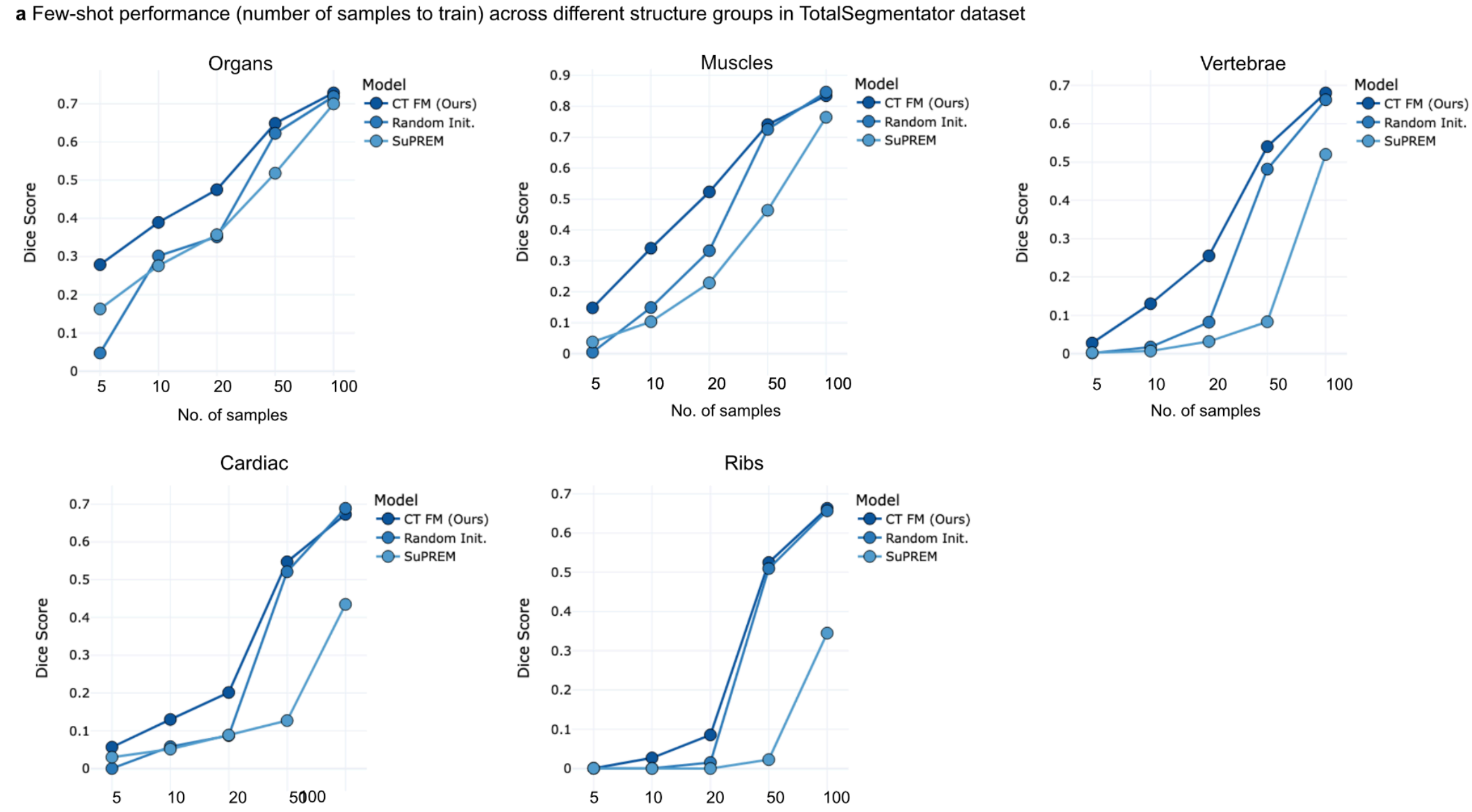

**Extended Data Figure 2:** Few-shot performance considering structure groups across the TotalSegmentator dataset

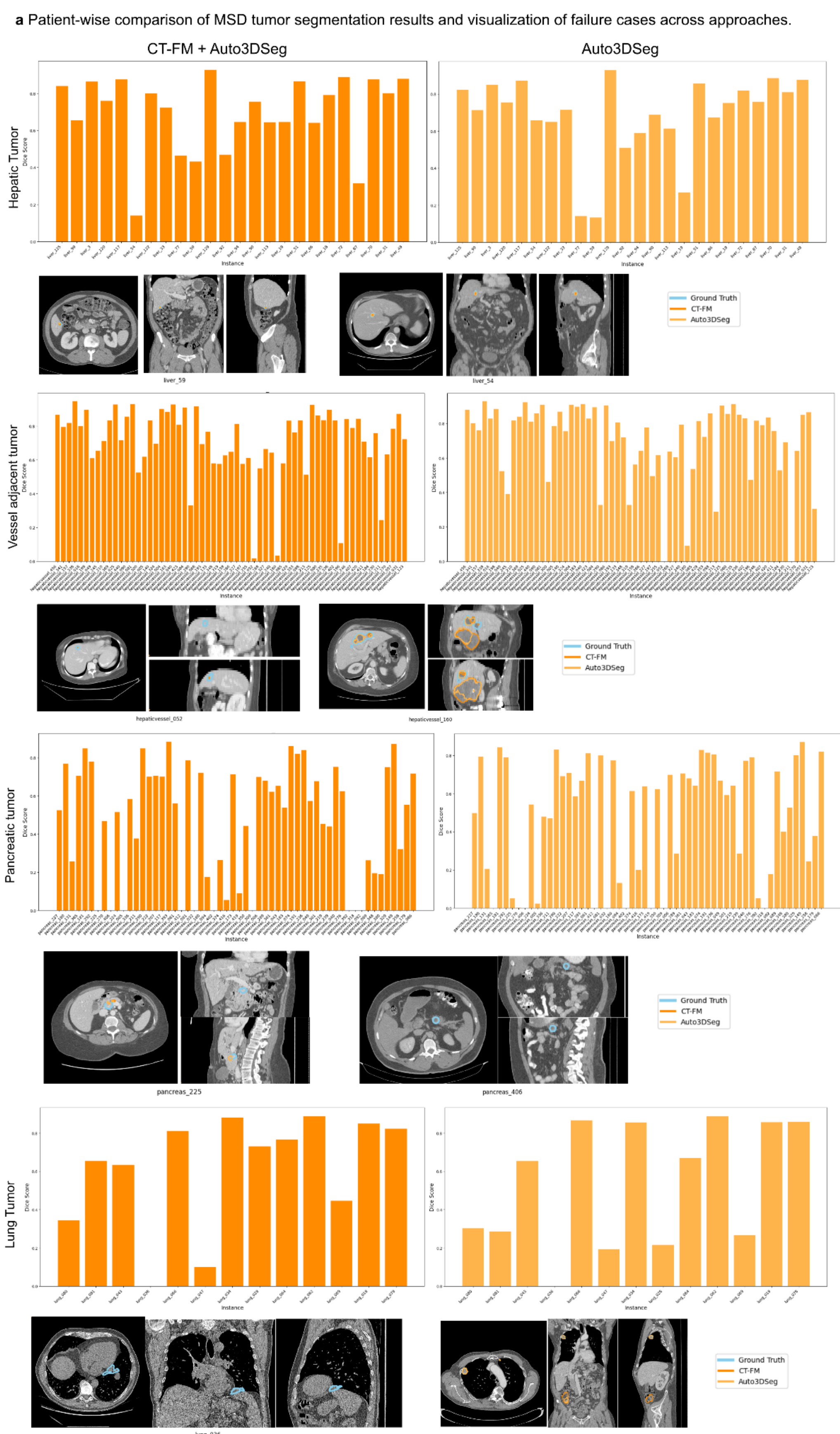

**Extended Data Figure 3:** Detailed patient-wise breakdown of comparisons on the MSD dataset with failure cases visualized.

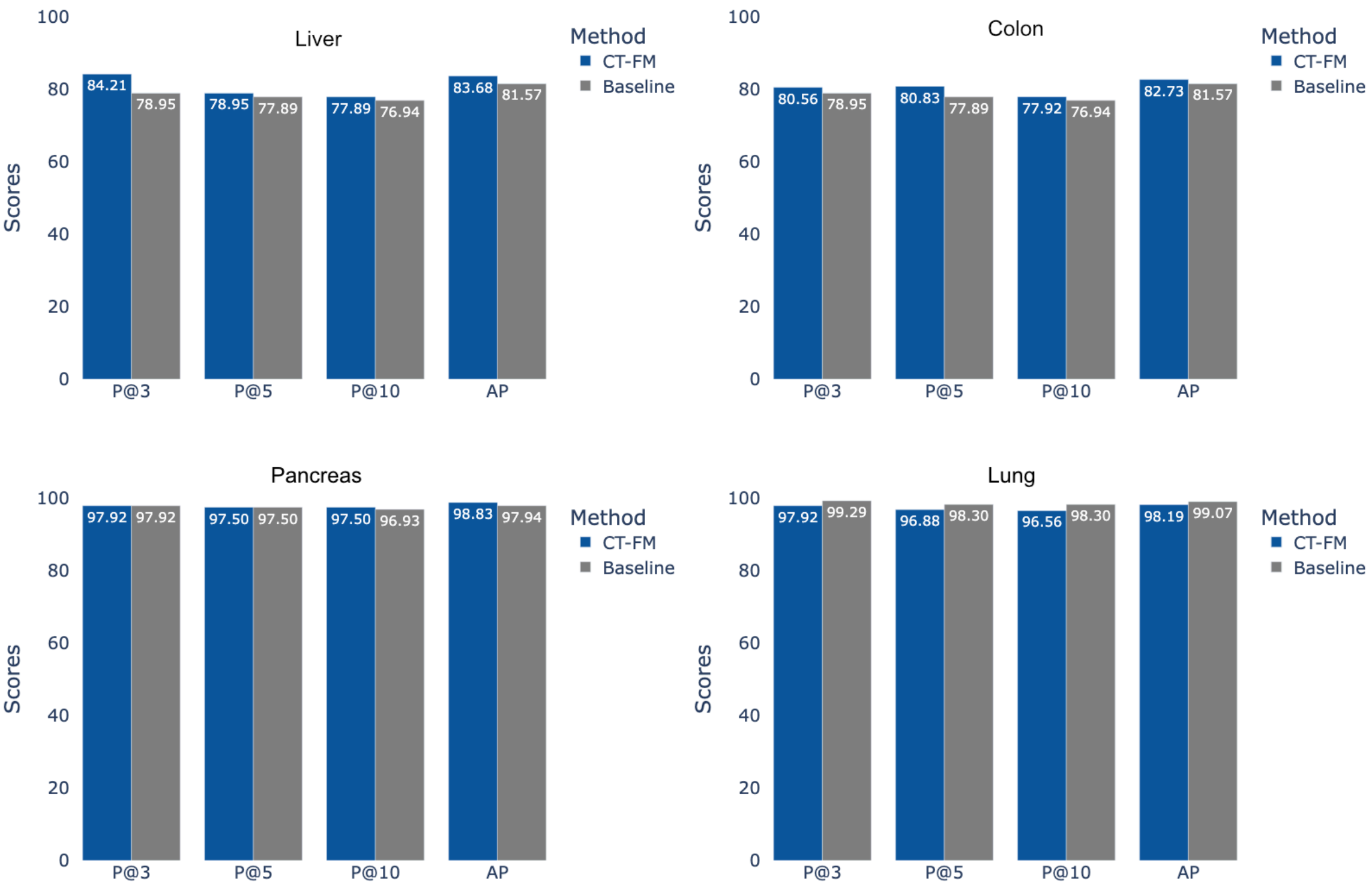

**Extended Data Figure 4:** Content-based retrieval detailed results for anatomical regions

**a** Comparison of anatomical clustering of CT-FM with baseline SuPREM across different values of perplexity

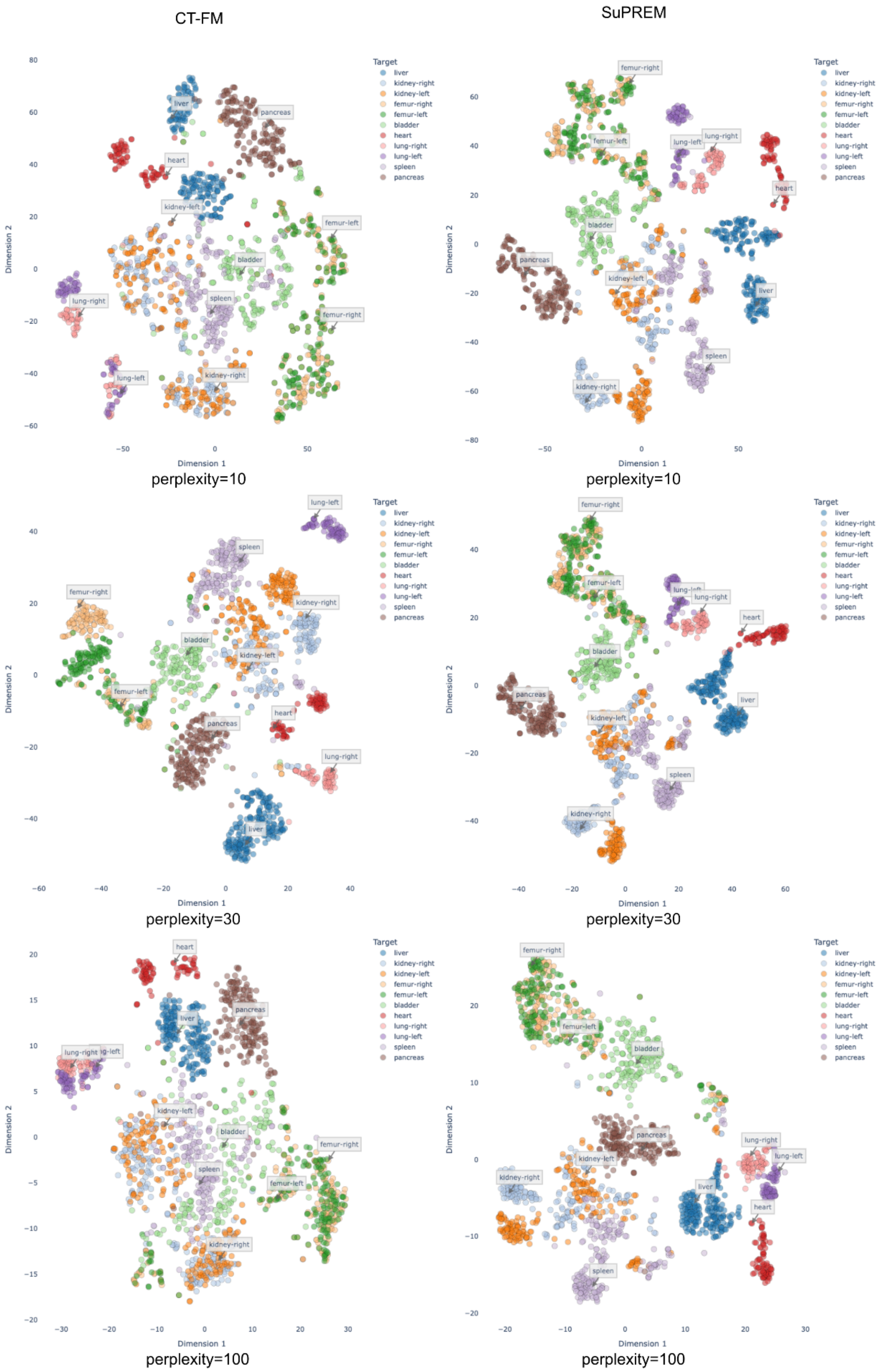

**Extended Data Figure 5:** Comparison of anatomical clustering between CT-FM and baseline

**a** Ablation of pre-training strategies evaluated on few-shot whole body segmentation on the TotalSegmentator dataset

**b** Ablation of number of crops on few-shot whole body segmentation

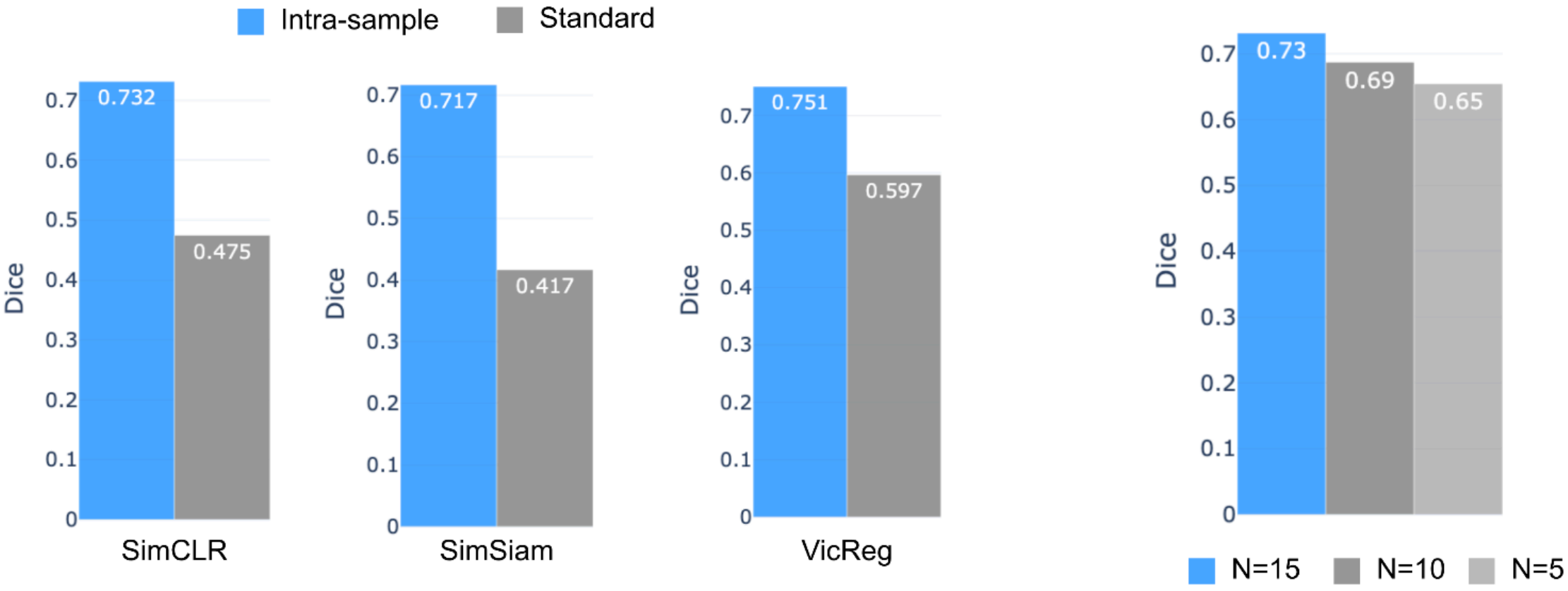

**c** Evolution of representation quality for whole body segmentation during the pre-training process.

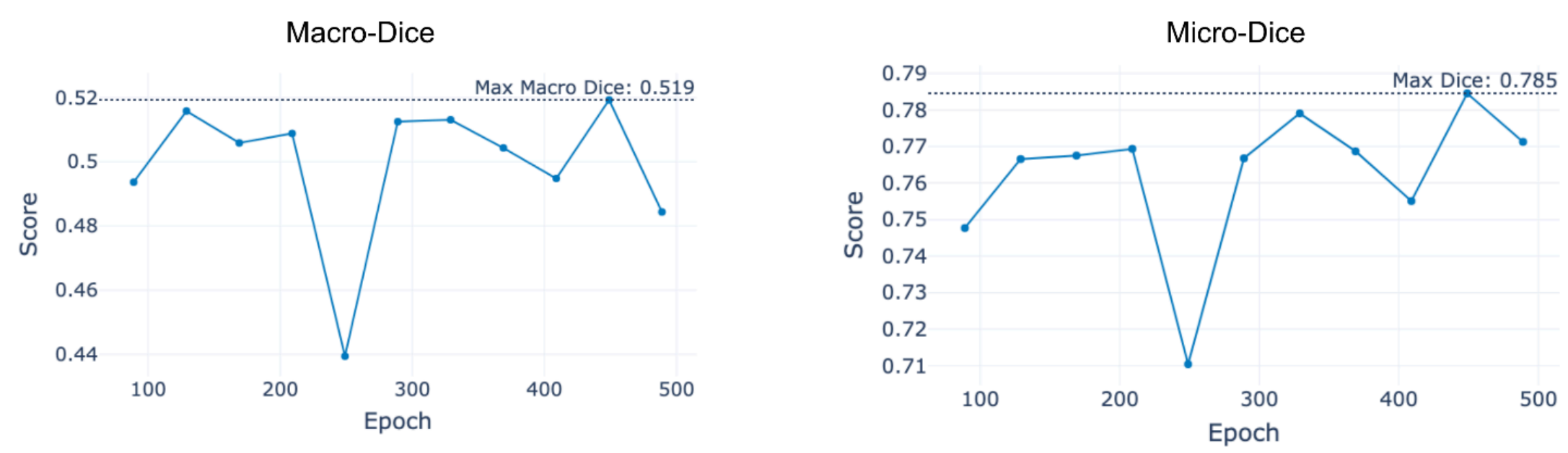

**d** Semantic concept matching between query patch and two selected target scans

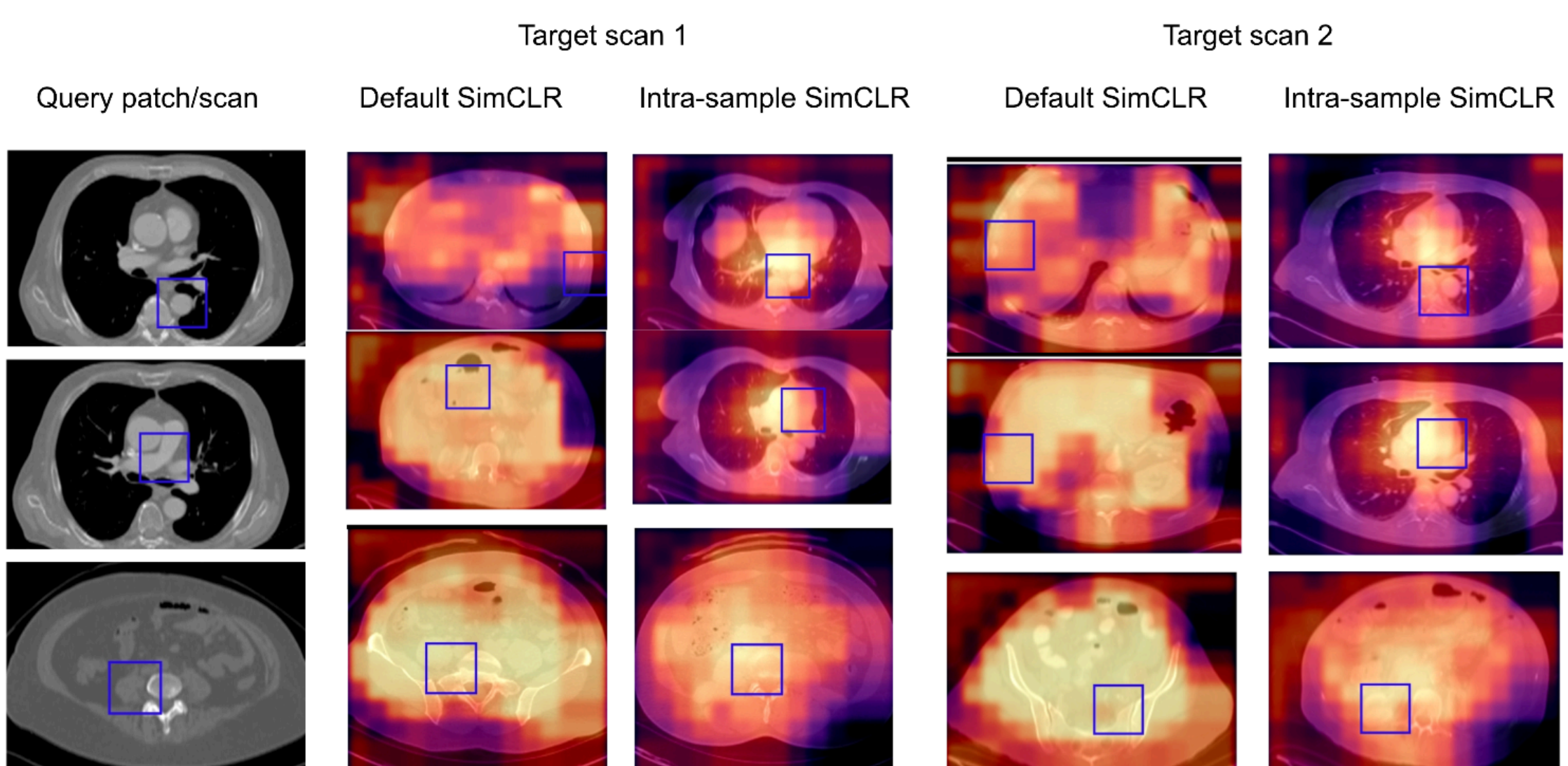

**Extended Data Figure 6:** Pre-training ablations and evolution of representation quality during pre-training

**a** Stability of patch embeddings on the RIDER test-retest dataset evaluated through cosine similarity

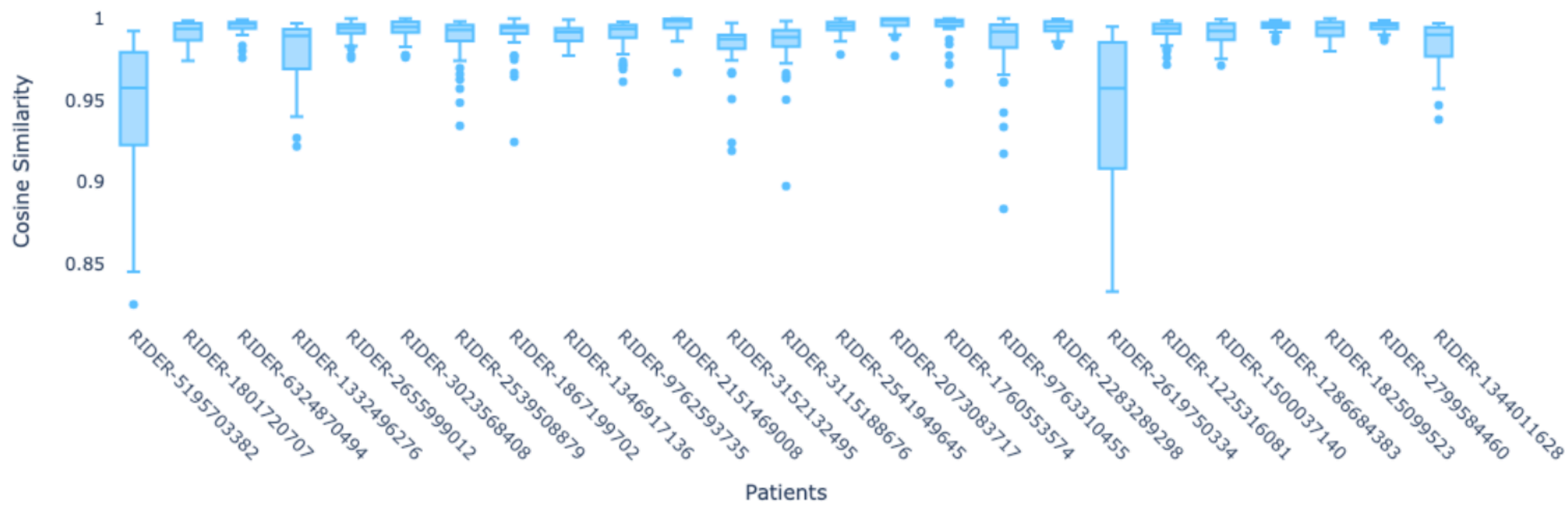

**b** Stability of patch embeddings on the RIDER test-retest dataset evaluated through mean squared error

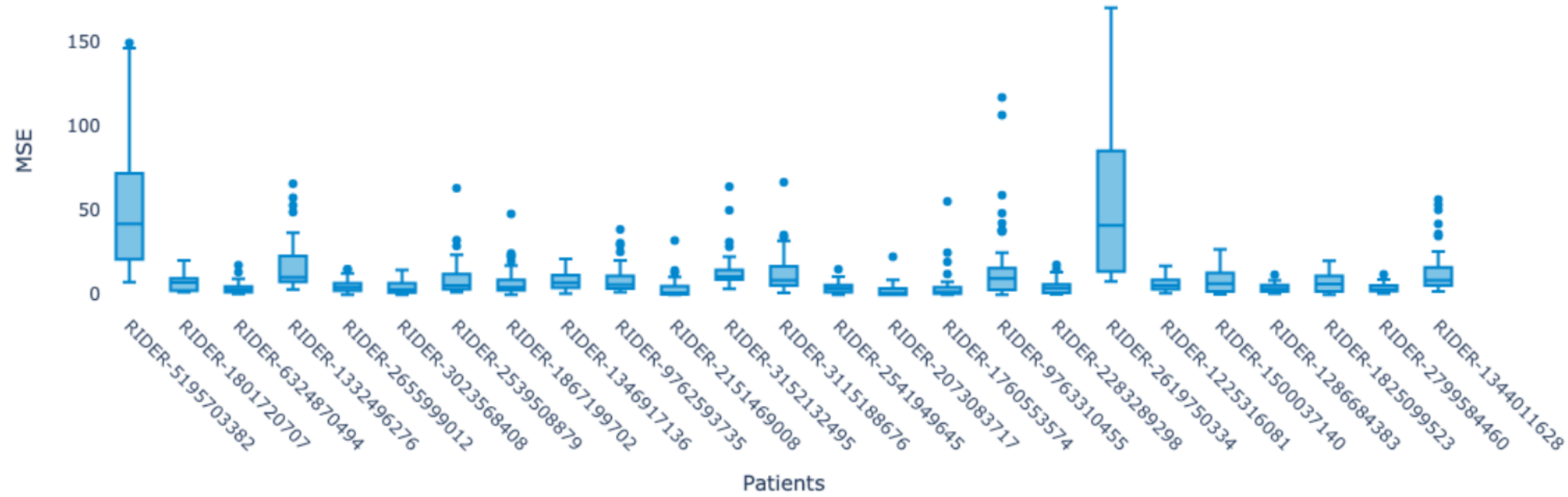

**Extended Data Figure 7:** Stability of CT-FM embeddings compared patch-wise between the RIDER dataset test-retest scans.

**a** Contrastive learning scenario for batch with limited variance between instances when compared to variance between views

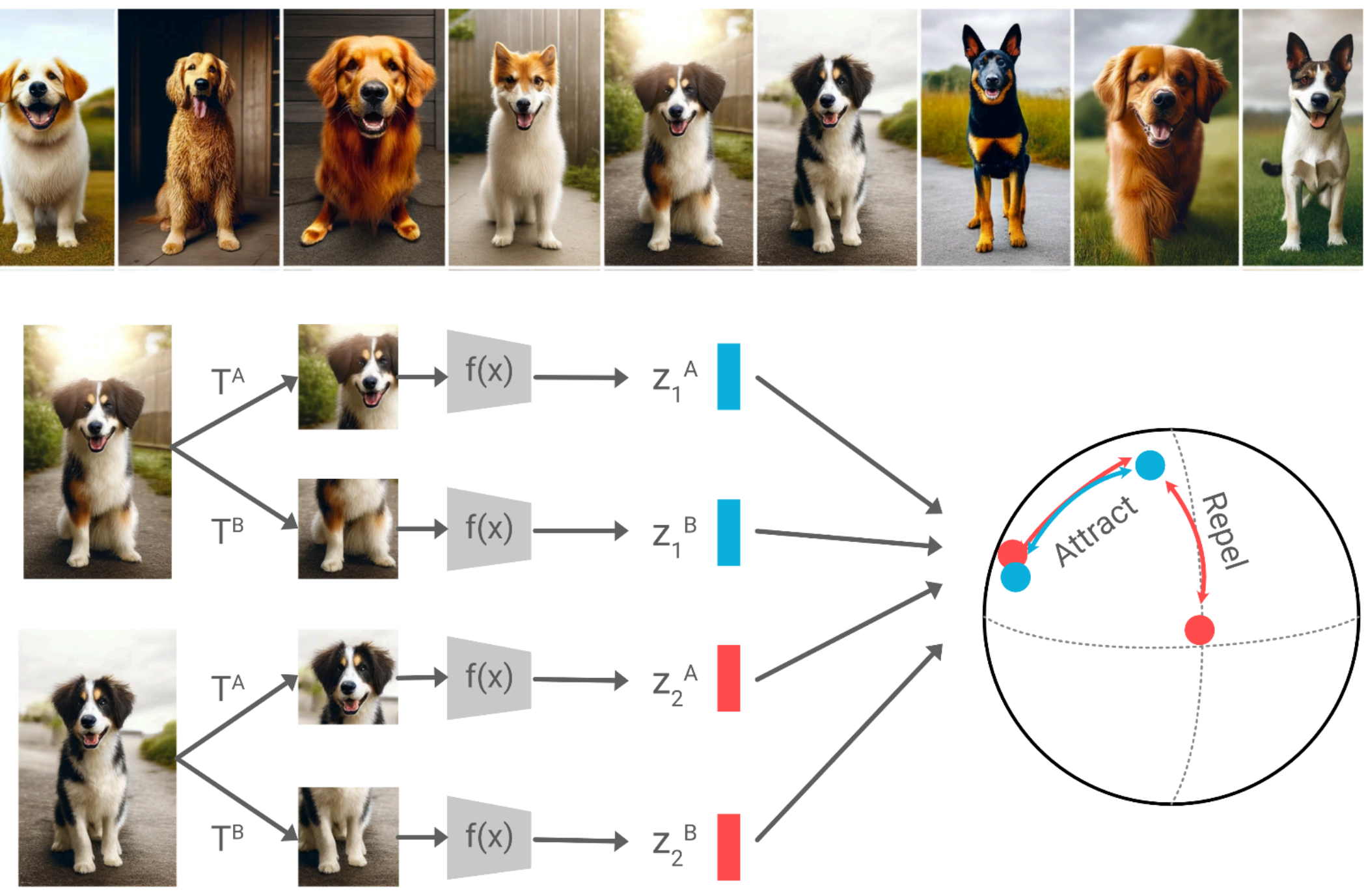

**b** Reformulation of the contrastive learning objective for limited variance medical imaging contexts

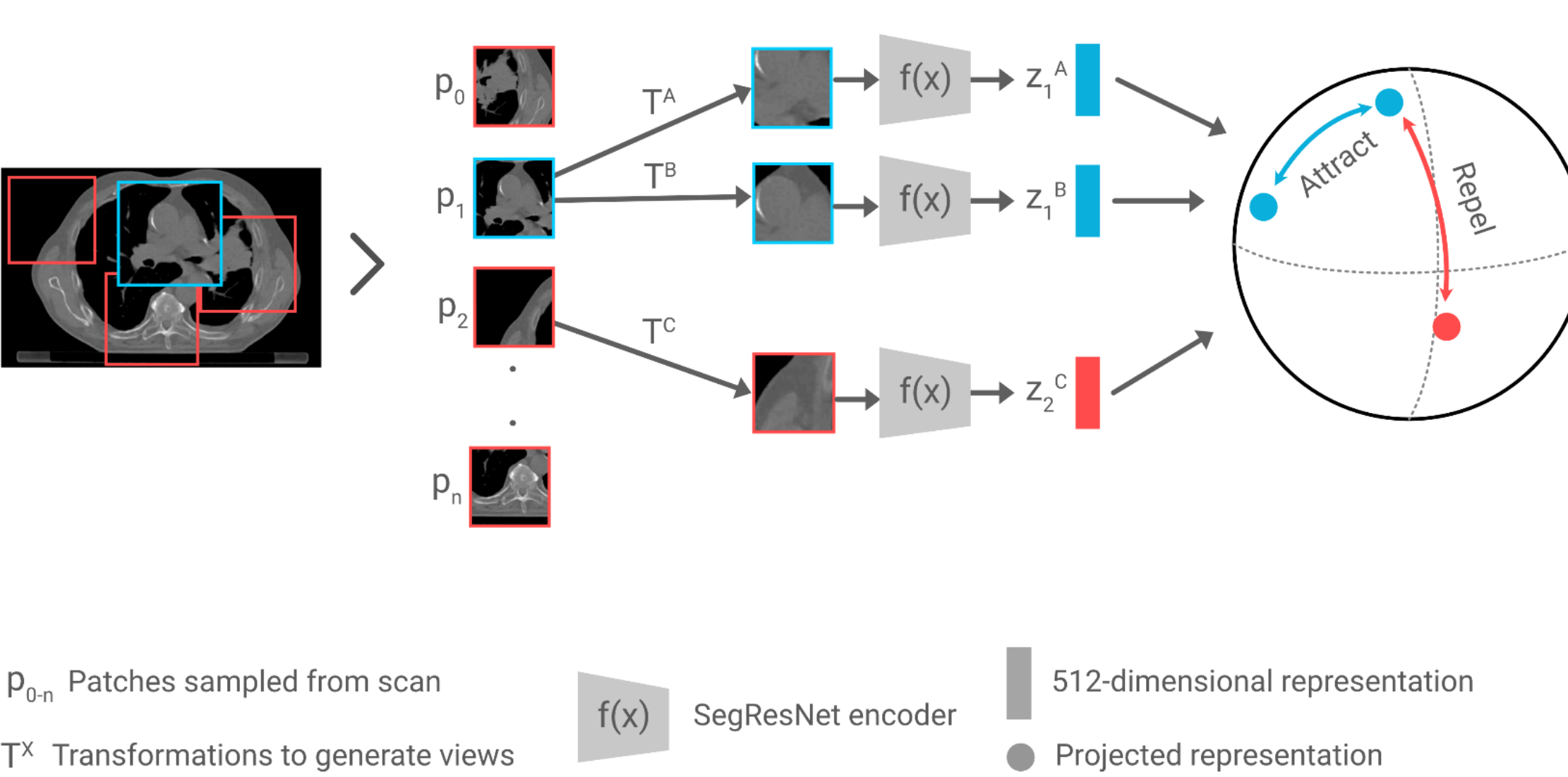

**Extended Data Figure 8:** Intra-sample pre-training adaptation for SimCLR

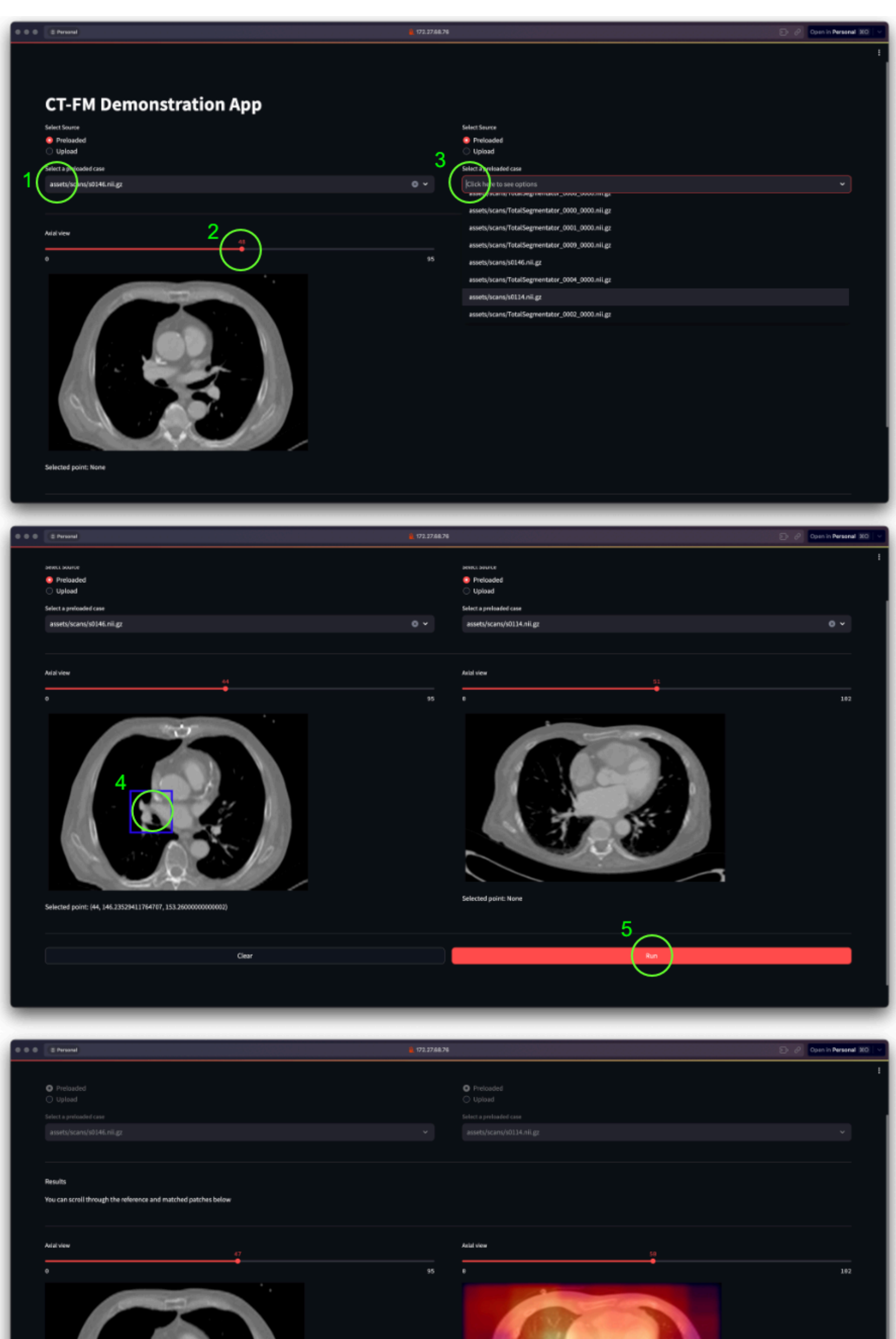

**Extended Data Figure 9:** Semantic search app workflow. Steps in the process are 1) select the source scan; 2) select the slice of interest; 3) select the target scan; 4) select a point around a region of interest in the source scan around which a 3D box will be cropped; 5) Run the semantic search framework to obtain the results

**Supplementary Information**

```sql
WITH
 idc_instances_per_series AS (
 SELECT
   SeriesInstanceUID,
   COUNT(DISTINCT(SOPInstanceUID)) AS num_instances,
   COUNT(DISTINCT(ARRAY_TO_STRING(ImagePositionPatient, "/"))) AS position_count,
   MAX(SAFE_CAST(SliceThickness AS float64)) AS max_SliceThickness,
   MIN(SAFE_CAST(SliceThickness AS float64)) AS min_SliceThickness,
   STRING_AGG(DISTINCT(SAFE_CAST("LOCALIZER" IN UNNEST(ImageType) AS string)), "") AS
has_localizer
 FROM
   `bigquery-public-data.idc_v14.dicom_all`
 WHERE
   Modality = "CT" and
   access = "Public"
 GROUP BY
   SeriesInstanceUID)
SELECT
 dicom_all.SeriesInstanceUID,
 ANY_VALUE(dicom_all.collection_id) as collection_id,
 ANY_VALUE(dicom_all.PatientID) AS PatientID,
 ANY_VALUE(idc_instances_per_series.num_instances) AS num_instances,
 ANY_VALUE(CONCAT("https://viewer.imaging.datacommons.cancer.gov/viewer/",
dicom_all.StudyInstanceUID, "?seriesInstanceUID=", dicom_all.SeriesInstanceUID)) AS
idc_url
FROM
 `bigquery-public-data.idc_v18.dicom_all` AS dicom_all
JOIN
 idc_instances_per_series
ON
 dicom_all.SeriesInstanceUID = idc_instances_per_series.SeriesInstanceUID
WHERE
 idc_instances_per_series.min_SliceThickness >= 1
 AND idc_instances_per_series.max_SliceThickness <= 5
 AND idc_instances_per_series.num_instances > 50
 AND idc_instances_per_series.num_instances/idc_instances_per_series.position_count = 1
 AND has_localizer = "false"
GROUP BY
 SeriesInstanceUID
```

S1: SQL Query for filtering pre-training dataset from the Imaging Data Commons

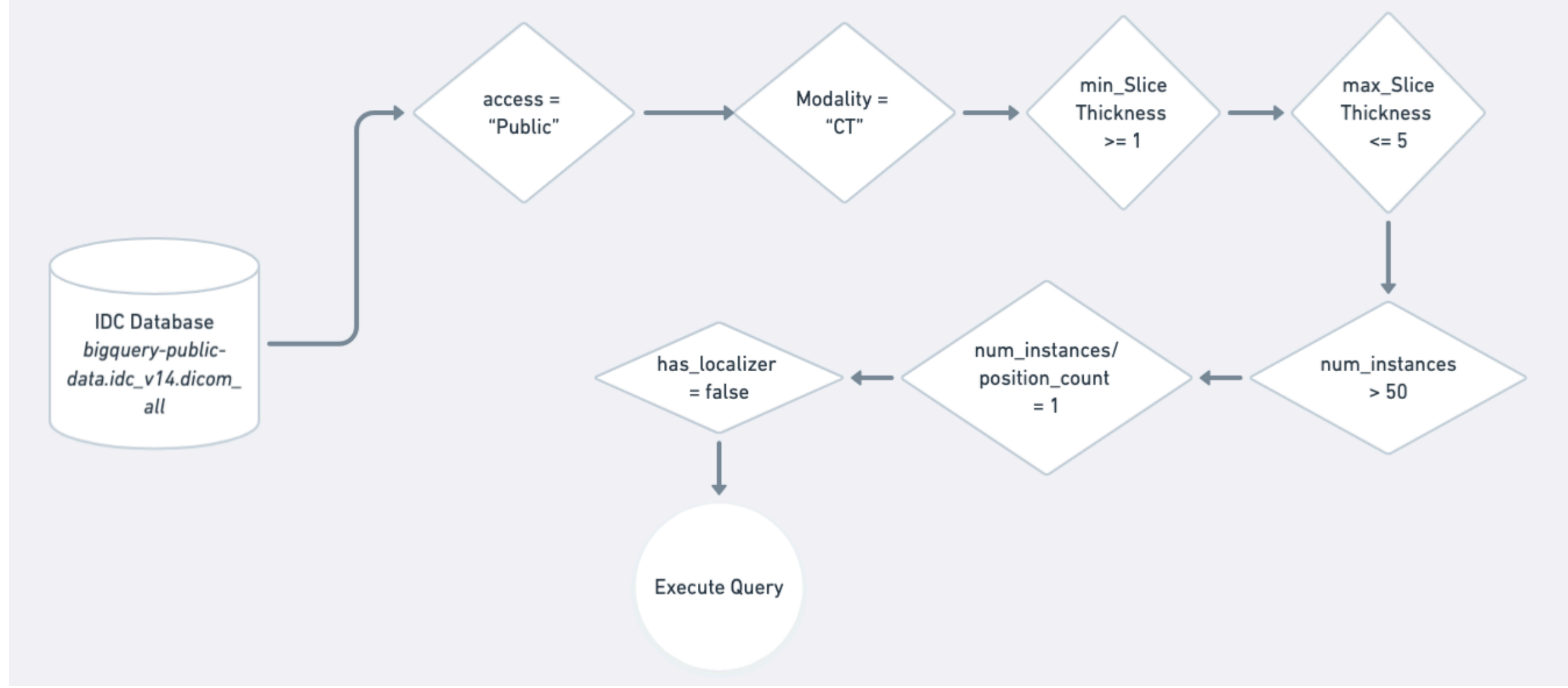

S2: Flowchart for filtering pre-training dataset from the Imaging Data Commons

| Label | CT-FM + Auto3DSeg | Auto3DSeg (our impl.) | Auto3DSeg | nnUNet | VISTA3D-auto |
|---|---|---|---|---|---|
| **Hepatic Tumor** | 0.695 | 0.681 | 0.616 | 0.617 | 0.588 |
| **Lung Tumor** | 0.610 | 0.532 | 0.562 | 0.554 | 0.614 |
| **Pancreatic Tumor** | 0.475 | 0.481 | 0.485 | 0.488 | 0.324 |
| **Hepatic (Vessel) Tumor** | 0.710 | 0.698 | 0.683 | 0.659 | 0.682 |

S3: Extended Comparison of CT-FM against other approaches on MSD dataset

| | Auto3dSeg | nnUNet | TotalSegmentator | VISTA3D auto | VISTA3D point | VISTA3D auto+point | CT FM (Ours) |
|---|---|---|---|---|---|---|---|
| spleen | 0.957 | 0.969 | 0.982 | 0.967 | 0.965 | 0.971 | 0.975 |
| right kidney | 0.949 | 0.94 | 0.962 | 0.934 | 0.93 | 0.948 | 0.943 |
| left kidney | 0.942 | 0.922 | 0.961 | 0.92 | 0.921 | 0.941 | 0.936 |
| gallbladder | 0.807 | 0.843 | 0.896 | 0.827 | 0.782 | 0.833 | 0.801 |
| liver | 0.964 | 0.965 | 0.982 | 0.968 | 0.944 | 0.974 | 0.971 |
| stomach | 0.929 | 0.935 | 0.96 | 0.931 | 0.917 | 0.939 | 0.941 |
| aorta | 0.954 | 0.961 | 0.961 | 0.959 | 0.949 | 0.965 | 0.968 |
| inferior vena cava | 0.892 | 0.902 | 0.896 | 0.883 | 0.695 | 0.896 | 0.919 |
| portal and splenic vein | 0.757 | 0.83 | 0.835 | 0.801 | 0.744 | 0.818 | 0.803 |
| pancreas | 0.845 | 0.856 | 0.917 | 0.86 | 0.833 | 0.877 | 0.865 |
| right adrenal gland | 0.805 | 0.877 | 0.909 | 0.863 | 0.834 | 0.869 | 0.846 |
| left adrenal gland | 0.808 | 0.866 | 0.914 | 0.873 | 0.851 | 0.881 | 0.853 |
| left lung upper lobe | 0.943 | 0.939 | 0.979 | 0.953 | 0.931 | 0.955 | 0.963 |
| left lung lower lobe | 0.928 | 0.953 | 0.964 | 0.938 | 0.899 | 0.944 | 0.946 |

| right lung upper lobe | 0.896 | 0.938 | 0.952 | 0.918 | 0.872 | 0.905 | 0.875 |
|---|---|---|---|---|---|---|---|
| right lung middle lobe | 0.905 | 0.939 | 0.952 | 0.916 | 0.909 | 0.93 | 0.929 |
| right lung lower lobe | 0.928 | 0.95 | 0.974 | 0.943 | 0.893 | 0.951 | 0.950 |
| vertebrae L5 | 0.909 | 0.93 | 0.946 | 0.916 | 0.916 | 0.933 | 0.926 |
| vertebrae L4 | 0.899 | 0.929 | 0.947 | 0.899 | 0.917 | 0.933 | 0.914 |
| vertebrae L3 | 0.892 | 0.927 | 0.967 | 0.925 | 0.934 | 0.957 | 0.918 |
| vertebrae L2 | 0.925 | 0.928 | 0.975 | 0.936 | 0.95 | 0.968 | 0.936 |
| vertebrae L1 | 0.904 | 0.917 | 0.964 | 0.919 | 0.934 | 0.955 | 0.919 |
| vertebrae T12 | 0.902 | 0.932 | 0.961 | 0.902 | 0.93 | 0.952 | 0.915 |
| vertebrae T11 | 0.899 | 0.922 | 0.97 | 0.9 | 0.93 | 0.952 | 0.912 |
| vertebrae T10 | 0.9 | 0.918 | 0.972 | 0.901 | 0.937 | 0.955 | 0.905 |
| vertebrae T9 | 0.886 | 0.918 | 0.976 | 0.901 | 0.936 | 0.96 | 0.896 |
| vertebrae T8 | 0.882 | 0.893 | 0.967 | 0.872 | 0.913 | 0.949 | 0.869 |
| vertebrae T7 | 0.822 | 0.886 | 0.92 | 0.831 | 0.89 | 0.92 | 0.829 |
| vertebrae T6 | 0.84 | 0.902 | 0.943 | 0.878 | 0.91 | 0.933 | 0.865 |
| vertebrae T5 | 0.869 | 0.923 | 0.944 | 0.891 | 0.904 | 0.93 | 0.888 |
| vertebrae T4 | 0.876 | 0.91 | 0.948 | 0.887 | 0.91 | 0.935 | 0.898 |
| vertebrae T3 | 0.888 | 0.926 | 0.95 | 0.895 | 0.903 | 0.935 | 0.908 |
| vertebrae T2 | 0.909 | 0.918 | 0.967 | 0.92 | 0.922 | 0.949 | 0.925 |
| vertebrae T1 | 0.907 | 0.945 | 0.969 | 0.933 | 0.926 | 0.95 | 0.939 |
| vertebrae C7 | 0.894 | 0.943 | 0.964 | 0.923 | 0.901 | 0.937 | 0.920 |
| vertebrae C6 | 0.839 | 0.84 | 0.941 | 0.852 | 0.864 | 0.912 | 0.874 |
| vertebrae C5 | 0.797 | 0.832 | 0.915 | 0.825 | 0.822 | 0.862 | 0.845 |
| vertebrae C4 | 0.86 | 0.859 | 0.944 | 0.904 | 0.881 | 0.917 | 0.890 |
| vertebrae C3 | 0.857 | 0.936 | 0.956 | 0.905 | 0.905 | 0.926 | 0.904 |
| vertebrae C2 | 0.908 | 0.953 | 0.972 | 0.91 | 0.872 | 0.933 | 0.914 |
| vertebrae C1 | 0.884 | 0.862 | 0.935 | 0.894 | 0.848 | 0.896 | 0.868 |
| esophagus | 0.874 | 0.913 | 0.952 | 0.907 | 0.886 | 0.916 | 0.910 |
| trachea | 0.926 | 0.945 | 0.974 | 0.941 | 0.91 | 0.946 | 0.949 |
| brain | 0.87 | 0.946 | 0.943 | 0.894 | 0.892 | 0.903 | 0.894 |
| left iliac artery | 0.822 | 0.896 | 0.916 | 0.895 | 0.872 | 0.906 | 0.882 |
| right iliac artery | 0.82 | 0.879 | 0.915 | 0.875 | 0.877 | 0.899 | 0.881 |
| left iliac vena | 0.841 | 0.898 | 0.941 | 0.917 | 0.899 | 0.925 | 0.910 |
| right iliac vena | 0.834 | 0.884 | 0.919 | 0.89 | 0.846 | 0.908 | 0.885 |

| small bowel | 0.854 | 0.868 | 0.948 | 0.821 | 0.846 | 0.869 | 0.862 |
| --- | --- | --- | --- | --- | --- | --- | --- |
| duodenum | 0.779 | 0.805 | 0.9 | 0.822 | 0.869 | 0.901 | 0.828 |
| colon | 0.882 | 0.882 | 0.948 | 0.898 | 0.819 | 0.906 | 0.908 |
| left rib 1 | 0.914 | 0.938 | 0.948 | 0.909 | 0.875 | 0.918 | 0.911 |
| right rib 1 | 0.934 | 0.927 | 0.966 | 0.932 | 0.909 | 0.943 | 0.914 |
| left rib 2 | 0.906 | 0.929 | 0.95 | 0.91 | 0.885 | 0.907 | 0.936 |
| right rib 2 | 0.908 | 0.936 | 0.947 | 0.903 | 0.887 | 0.927 | 0.931 |
| left rib 3 | 0.878 | 0.895 | 0.933 | 0.889 | 0.889 | 0.928 | 0.919 |
| right rib 3 | 0.865 | 0.912 | 0.925 | 0.866 | 0.884 | 0.916 | 0.906 |
| left rib 6 | 0.885 | 0.907 | 0.942 | 0.877 | 0.901 | 0.934 | 0.879 |
| right rib 6 | 0.902 | 0.888 | 0.955 | 0.89 | 0.91 | 0.941 | 0.894 |
| left rib 9 | 0.91 | 0.901 | 0.953 | 0.897 | 0.916 | 0.937 | 0.900 |
| right rib 9 | 0.911 | 0.883 | 0.949 | 0.893 | 0.906 | 0.938 | 0.886 |
| left rib 10 | 0.891 | 0.904 | 0.949 | 0.903 | 0.911 | 0.938 | 0.891 |
| right rib 10 | 0.885 | 0.873 | 0.912 | 0.883 | 0.871 | 0.909 | 0.907 |
| left rib 11 | 0.905 | 0.938 | 0.945 | 0.907 | 0.875 | 0.912 | 0.901 |
| right rib 11 | 0.933 | 0.946 | 0.959 | 0.924 | 0.888 | 0.929 | 0.891 |
| left rib 12 | 0.906 | 0.938 | 0.931 | 0.891 | 0.854 | 0.9 | 0.883 |
| right rib 12 | 0.928 | 0.942 | 0.949 | 0.906 | 0.882 | 0.926 | 0.894 |
| right rib 4 | 0.905 | 0.893 | 0.916 | 0.876 | 0.877 | 0.914 | 0.911 |

| right rib 5 | 0.9 | 0.929 | 0.951 | 0.886 | 0.907 | 0.932 | 0.874 |
|---|---|---|---|---|---|---|---|
| right rib 7 | 0.903 | 0.914 | 0.96 | 0.884 | 0.915 | 0.942 | 0.899 |
| right rib 8 | 0.888 | 0.928 | 0.959 | 0.887 | 0.913 | 0.941 | 0.886 |
| left humerus | 0.911 | 0.867 | 0.93 | 0.854 | 0.881 | 0.903 | 0.902 |
| right humerus | 0.916 | 0.794 | 0.94 | 0.873 | 0.884 | 0.913 | 0.917 |
| left scapula | 0.91 | 0.949 | 0.959 | 0.911 | 0.887 | 0.921 | 0.931 |
| right scapula | 0.916 | 0.923 | 0.959 | 0.922 | 0.887 | 0.92 | 0.937 |
| left clavicula | 0.955 | 0.917 | 0.975 | 0.952 | 0.931 | 0.956 | 0.953 |
| right clavicula | 0.937 | 0.94 | 0.973 | 0.945 | 0.933 | 0.952 | 0.953 |
| left femur | 0.944 | 0.882 | 0.97 | 0.94 | 0.944 | 0.954 | 0.960 |
| right femur | 0.944 | 0.911 | 0.98 | 0.945 | 0.957 | 0.959 | 0.971 |
| left hip | 0.944 | 0.937 | 0.975 | 0.947 | 0.938 | 0.955 | 0.975 |
| right hip | 0.939 | 0.932 | 0.986 | 0.95 | 0.961 | 0.959 | 0.970 |
| sacrum | 0.925 | 0.933 | 0.958 | 0.914 | 0.925 | 0.922 | 0.942 |
| left gluteus maximus | 0.923 | 0.927 | 0.972 | 0.94 | 0.938 | 0.949 | 0.961 |
| right gluteus maximus | 0.917 | 0.93 | 0.978 | 0.937 | 0.937 | 0.949 | 0.959 |

| left gluteus medius | 0.919 | 0.926 | 0.973 | 0.931 | 0.923 | 0.923 | 0.947 |
|---|---|---|---|---|---|---|---|
| right gluteus medius | 0.908 | 0.927 | 0.978 | 0.938 | 0.937 | 0.946 | 0.958 |
| left gluteus minimus | 0.875 | 0.917 | 0.965 | 0.914 | 0.903 | 0.919 | 0.933 |
| right gluteus minimus | 0.876 | 0.92 | 0.967 | 0.915 | 0.896 | 0.921 | 0.935 |
| left autochthon | 0.939 | 0.934 | 0.978 | 0.951 | 0.932 | 0.953 | 0.964 |
| right autochthon | 0.941 | 0.932 | 0.976 | 0.941 | 0.927 | 0.947 | 0.961 |
| left iliopsoas | 0.876 | 0.91 | 0.965 | 0.921 | 0.898 | 0.926 | 0.934 |
| right iliopsoas | 0.876 | 0.916 | 0.952 | 0.907 | 0.898 | 0.914 | 0.925 |
| bladder | 0.89 | 0.906 | 0.934 | 0.899 | 0.895 | 0.915 | 0.916 |
| left atrial appendage | 0.864 | 0.906 | 0.942 | 0.901 | 0.873 | 0.91 | 0.905 |
| brachiocephalic trunk | 0.872 | 0.899 | 0.936 | 0.892 | 0.858 | 0.915 | 0.911 |
| left brachiocephalic vein | 0.881 | 0.919 | 0.942 | 0.904 | 0.885 | 0.898 | 0.919 |
| right brachiocephali | 0.862 | 0.909 | 0.922 | 0.884 | 0.869 | 0.901 | 0.890 |

| c vein | | | | | | | |
|---|---|---|---|---|---|---|---|
| left common carotid artery | 0.826 | 0.884 | 0.925 | 0.868 | 0.828 | 0.891 | 0.879 |
| right common carotid artery | 0.755 | 0.858 | 0.885 | 0.811 | 0.784 | 0.844 | 0.818 |
| costal cartilages | 0.844 | 0.868 | 0.888 | 0.856 | 0.833 | 0.864 | 0.862 |
| heart | 0.932 | 0.928 | 0.937 | 0.919 | 0.916 | 0.924 | 0.936 |
| left kidney cyst | 0.623 | 0.858 | 0.892 | 0.618 | 0.752 | 0.858 | 0.449 |
| right kidney cyst | 0.568 | 0.841 | 0.716 | 0.606 | 0.615 | 0.681 | 0.277 |
| prostate | 0.743 | 0.752 | 0.808 | 0.744 | 0.745 | 0.774 | 0.759 |
| pulmonary vein | 0.838 | 0.82 | 0.916 | 0.83 | 0.747 | 0.863 | 0.877 |
| skull | 0.909 | 0.849 | 0.893 | 0.827 | 0.769 | 0.857 | 0.871 |
| spinal cord | 0.911 | 0.95 | 0.959 | 0.934 | 0.905 | 0.911 | 0.936 |
| sternum | 0.896 | 0.906 | 0.897 | 0.899 | 0.884 | 0.911 | 0.920 |
| left subclavian artery | 0.833 | 0.901 | 0.929 | 0.877 | 0.857 | 0.892 | 0.891 |
| right subclavian artery | 0.818 | 0.87 | 0.916 | 0.861 | 0.85 | 0.885 | 0.875 |
| superior vena | 0.894 | 0.899 | 0.932 | 0.888 | 0.905 | 0.923 | 0.932 |
| cava | | | | | | | |
| thyroid gland | 0.832 | 0.886 | 0.908 | 0.866 | 0.853 | 0.89 | 0.886 |
| vertebrae S1 | 0.87 | 0.906 | 0.925 | 0.89 | 0.88 | 0.909 | 0.915 |

S4: Extended Comparison of CT-FM against other approaches on the TotalSegmentatorV2 dataset

| Collection ID | Number of Scans | Description |
| --- | --- | --- |
| nlst | 128037 | The National Lung Screening Trial (NLST) dataset includes low-dose CT scans from a large, randomized trial evaluating lung cancer screening. |
| phantom_fda | 3576 | This collection contains CT scans of a Catphan phantom, used for quality control and standardization in imaging. |
| covid_19_ny_sbu | 1753 | The COVID-19 CT scan dataset from Stony Brook University, containing images of patients diagnosed with COVID-19. |
| ct_colonography | 1711 | This collection includes CT colonography scans, a minimally invasive method for colorectal cancer screening. |
| acrin_nsclc_fdg_pet | 1283 | The ACRIN NSCLC FDG-PET dataset contains both CT and PET scans of patients with non-small cell lung cancer from a clinical trial. |
| lidc_idri | 974 | The Lung Image Database Consortium (LIDC) and Image Database Resource Initiative (IDRI) dataset includes lung CT scans with annotations by multiple radiologists. |
| 4d_lung | 830 | This collection includes 4D (time-resolved) CT scans of the lung, capturing respiratory |

| | | motion. |
|---|---|---|
| tcga_kirc | 812 | The Cancer Genome Atlas (TCGA) Kidney Renal Clear Cell Carcinoma (KIRC) dataset contains CT and other images of patients with renal cell carcinoma. |
| lung_pet_ct_dx | 546 | This collection contains PET/CT scans of patients with suspected or confirmed lung cancer. |
| hcc_tace_seg | 477 | This collection includes CT scans of liver tumors from patients undergoing Transarterial Chemoembolization (TACE), with segmentation masks. |
| nsclc_radiogenomics | 427 | This dataset combines CT imaging with genomic data from patients with non-small cell lung cancer. |
| nsclc_radiomics | 421 | This collection of CT scans is related to radiomics analysis of patients with non-small cell lung cancer. |
| tcga_blca | 409 | The TCGA Bladder Urothelial Carcinoma (BLCA) dataset contains CT and other images from patients with bladder cancer. |
| cptac_ucec | 393 | The Clinical Proteomic Tumor Analysis Consortium (CPTAC) Uterine Corpus Endometrial Carcinoma (UCEC) dataset combines imaging with proteomic data. |

| tcga_ov | 384 | The TCGA Ovarian Cancer (OV) dataset includes CT and other images from patients with ovarian cancer. |
|---|---|---|
| c4kc_kits | 366 | The Cancer Imaging Archive (TCIA) "C4KC" Kits collection includes phantom and clinical images for quantitative imaging analysis. |
| cptac_ccrcc | 333 | The CPTAC Clear Cell Renal Cell Carcinoma (CCRCC) dataset includes CT and other imaging data linked to proteomic profiles. |
| tcga_ucec | 330 | The TCGA Uterine Corpus Endometrial Carcinoma (UCEC) dataset includes imaging data linked to genomic and clinical data. |
| cptac_pda | 305 | The CPTAC Pancreatic Ductal Adenocarcinoma (PDA) dataset combines proteomic and imaging data for pancreatic cancer patients. |
| pediatric_ct_seg | 296 | This collection contains pediatric CT scans with segmentation masks for various anatomical structures. |
| tcga_lihc | 284 | The TCGA Liver Hepatocellular Carcinoma (LIHC) dataset includes imaging and associated genomic data for patients with liver cancer. |
| acrin_flt_breast | 279 | The ACRIN [1]8F-FLT Breast dataset contains PET/CT scans of breast |

| | | cancer patients with additional [18F]-FLT tracer. |
|---|---|---|
| anti_pd_1_lung | 265 | This dataset includes CT scans from lung cancer patients treated with anti-PD-1 immunotherapy. |
| cmb_crc | 251 | The Cancer Moonshot Biobank (CMB) Colorectal Cancer (CRC) dataset includes CT scans of colorectal cancer patients with associated clinical data. |
| tcga_stad | 237 | The TCGA Stomach Adenocarcinoma (STAD) dataset includes imaging data linked to genomic information for stomach cancer. |
| rider_lung_pet_ct | 235 | The Radiological Image Database for Research (RIDER) Lung PET/CT dataset contains a collection of PET and CT scans of patients with lung cancer. |
| qiba_ct_1c | 211 | This is the Quantitative Imaging Biomarkers Alliance (QIBA) CT phantom study, containing CT scans from a multisite study. |
| pancreatic_ct_cbct_seg | 200 | This collection includes CT and Cone Beam CT (CBCT) scans of the pancreas with segmentation masks for various structures. |
| tcga_luad | 183 | The TCGA Lung Adenocarcinoma (LUAD) dataset contains CT and other images linked to genomic data from |

| | | patients with lung adenocarcinoma. |
|---|---|---|
| stageii_colorectal_ct | 181 | This collection includes CT scans of patients with stage II colorectal cancer. |
| midrc_ricord_1a | 163 | Part of the Medical Imaging Data Resource Center (MIDRC), this collection includes COVID-19 CT scans (Phase 1a) along with clinical annotations. |
| cptac_lscc | 159 | The CPTAC Lung Squamous Cell Carcinoma (LSCC) dataset includes CT imaging and proteomic data for lung squamous cell carcinoma patients. |
| tcga_lusc | 133 | The TCGA Lung Squamous Cell Carcinoma (LUSC) dataset includes CT and other images linked to genomic data for patients with lung squamous cell carcinoma. |
| cptac_luad | 133 | The CPTAC Lung Adenocarcinoma (LUAD) dataset combines imaging with proteomic data for lung adenocarcinoma patients. |
| cmb_mel | 124 | The Cancer Moonshot Biobank (CMB) Melanoma (MEL) dataset includes CT scans of melanoma patients. |
| midrc_ricord_1b | 118 | Part of the Medical Imaging Data Resource Center (MIDRC), this collection includes COVID-19 CT scans |

| | | (Phase 1b) along with clinical annotations. |
|---|---|---|
| pelvic_reference_data | 116 | This collection provides reference CT scans of the pelvic region. |
| covid_19_ar | 113 | This collection contains COVID-19 CT scans with associated metadata from patients in Argentina. |
| qin_breast | 110 | The Quantitative Imaging Network (QIN) Breast dataset contains imaging data from patients with breast cancer. |
| cmb_lca | 107 | The Cancer Moonshot Biobank (CMB) Lung Cancer Adenocarcinoma (LCA) dataset contains CT scans from lung cancer patients. |
| tcga_esca | 94 | The TCGA Esophageal Carcinoma (ESCA) dataset includes CT imaging data linked to genomic data from patients with esophageal cancer. |
| nsclc_radiomics_genomics | 88 | This dataset combines radiomics features extracted from CT scans with genomic data in non-small cell lung cancer. |
| naf_prostate | 74 | The dataset contains prostate MRIs from patients in a national active surveillance study. |
| ctpred_sunitinib_pannet | 73 | The dataset contains CT scans of patients with pancreatic neuroendocrine tumors (pNET) treated with |

| | | Sunitinib. |
|---|---|---|
| spie_aapm_lung_ct_challenge | 70 | This collection is a challenge dataset for lung CT image analysis, containing scans from a SPIE-AAPM challenge. |
| cptac_cm | 63 | The CPTAC Cutaneous Melanoma (CM) dataset includes imaging data linked to proteomic and genomic data from melanoma patients. |
| rider_lung_ct | 63 | The RIDER Lung CT dataset contains CT scans of lung cancer patients collected from the RIDER project. |
| cptac_sar | 63 | The CPTAC Sarcoma (SAR) dataset combines imaging data with proteomic profiles of patients with sarcoma. |
| lctsc | 60 | This collection contains a longitudinal CT dataset of lung cancer patients from the Lung Cancer Therapy Study Consortium (LCTSC). |
| ct_vs_pet_ventilation_imaging | 59 | This collection contains CT and PET scans for assessment of ventilation imaging. |
| lungct_diagnosis | 52 | This collection contains a variety of lung CT scans for the task of diagnosis classification. |
| soft_tissue_sarcoma | 51 | This collection consists of soft tissue sarcoma CT scans. |
| tcga_kirp | 47 | The TCGA Kidney Renal Papillary Cell Carcinoma (KIRP) dataset includes CT and other images for |

| | | patients with kidney papillary cell carcinoma. |
|---|---|---|
| qin_lung_ct | 39 | The Quantitative Imaging Network (QIN) Lung CT dataset contains scans from a clinical research study on lung cancer. |
| tcga_coad | 37 | The TCGA Colon Adenocarcinoma (COAD) dataset includes CT and other images linked to genomic data for colon cancer patients. |
| tcga_kich | 33 | The TCGA Kidney Chromophobe (KICH) dataset contains CT and other images linked to genomic data for kidney chromophobe carcinoma patients. |
| dro_toolkit | 32 | This collection provides some data and tools for use with the Data Release Orchestrator (DRO). |
| nsclc_radiomics_interobserver1 | 22 | This is a small dataset used to study interobserver variability in radiomics feature extraction from lung cancer CT scans. |
| cmb_mml | 21 | The Cancer Moonshot Biobank (CMB) Multiple Myeloma (MML) dataset includes CT scans of multiple myeloma patients. |
| breast_diagnosis | 18 | This collection is made of a small breast CT dataset for use in diagnosis tasks. |
| lung_fused_ct_pathology | 15 | This dataset contains CT scans of lung nodules matched to |

| | | histopathology data. |
|---|---|---|
| tcga_prad | 13 | The TCGA Prostate Adenocarcinoma (PRAD) dataset contains CT and other images linked to genomic data for prostate cancer patients. |
| cmb_pca | 12 | The Cancer Moonshot Biobank (CMB) Prostate Cancer (PCA) dataset includes CT scans of prostate cancer patients. |
| tcga_thca | 11 | The TCGA Thyroid Carcinoma (THCA) dataset includes CT and other images from patients with thyroid cancer. |
| cmb_gec | 8 | The Cancer Moonshot Biobank (CMB) Gastroesophageal Cancer (GEC) dataset includes CT scans of gastroesophageal cancer patients. |
| pseudo_phi_dicom_data | 4 | This collection contains pseudo-anonymized DICOM data for testing DICOM data handling. |
| tcga_read | 3 | The TCGA Rectum Adenocarcinoma (READ) dataset contains imaging and genomic data from patients with rectal cancer. |
| tcga_sarc | 3 | The TCGA Sarcoma (SARC) dataset includes CT and other images from patients with sarcoma. |
| lung_phantom | 1 | This collection contains CT scans of a lung phantom. |

S5: IDC collections used with the number of scans sourced from them and a short description of the collection.